\documentclass[10pt,twocolumn,aps,prxquantum,superscriptaddress,showpacs,tightenlines,pdflatex,longbibliography]{revtex4-2}

\usepackage{amsthm}
\usepackage{gensymb}
\usepackage{amsmath}            %serve per le subequazioni
\usepackage{amssymb}            %serve per il simbolo "marchio registrato", \circledR
\usepackage{mathtools}
\usepackage{graphicx}           %serve per le figure eps, ps etcyy
\usepackage{xcolor}
\usepackage{braket}
\usepackage{etoolbox}
\usepackage{epstopdf}
\usepackage{multirow}

\usepackage{xcolor}
\usepackage{bm}
\usepackage[normalem]{ulem}
\makeatletter
\let\cat@comma@active\@empty

\usepackage[colorlinks,citecolor=blue,urlcolor=blue,bookmarks=false,hypertexnames=true]{hyperref} 
\newcommand{\id}{\openone}

\newcommand{\comment}[1]{}

\theoremstyle{plain}

\begin{document}
\title{Charging efficiency bursts in a quantum battery with cyclic indefinite causal order}

\author{Po-Rong Lai}
%\email{poronglai@gmail.com}
\affiliation{Department of Physics, National Cheng Kung University, Tainan 701, Taiwan}
\affiliation{Center for Quantum Frontiers of Research and Technology, NCKU, Tainan 701, Taiwan}

\author{Hsien-Chao Jan}
\affiliation{Department of Chemistry, University of Rochester, Rochester, New York 14627, USA}

\author{Jhen-Dong Lin}
% \email{jhendonglin@gmail.com}
\affiliation{Department of Physics, National Cheng Kung University, Tainan 701, Taiwan}
\affiliation{Center for Quantum Frontiers of Research and Technology, NCKU, Tainan 701, Taiwan}

% \author{Yi-Te Huang}
% \affiliation{Department of Physics and Center for Quantum Frontiers of Research \&
% Technology (QFort), National Cheng Kung University, Tainan 701, Taiwan}

\author{Yueh-Nan Chen}
\email{yuehnan@mail.ncku.edu.tw}
\affiliation{Department of Physics, National Cheng Kung University, Tainan 701, Taiwan}
\affiliation{Center for Quantum Frontiers of Research and Technology, NCKU, Tainan 701, Taiwan}
\affiliation{Physics Division, National Center for Theoretical Sciences, Taipei 106319, Taiwan}

\begin{abstract}
Enhancement of quantum battery performance is a popular subject in quantum thermodynamics. An interesting phenomenon is the quick charging effect [Phys. Rev. Res. 6, 023136 (2024)], which has been explored by utilizing a quantum interferometric technique known as superposition of trajectories. A similar technique used to boost quantum battery performance is indefinite causal order. Here, we propose a new charging protocol that utilizes cyclic indefinite causal order, whereby $N$ charging sequences are superposed when utilizing $N$ chargers. We observe charging efficiency bursts when implementing our cyclic indefinite charging protocol. The duration of these bursts increase with $N$. Additionally, we present a circuit model to implement our charging protocol for the two-charger scenario and perform proof-of-concept demonstrations on IonQ, Quantinuum and IBMQ quantum processors. The results validate the existence of charging efficiency bursts as shown by our theoretical analysis and numerical simulations.

% However, current literature focuses on the simplest adoption of indefinite causal order, where just two charging sequences are superposed during implementation.
\end{abstract}

\maketitle

\section{Introduction}

Quantum batteries (QBs) are thermal devices that have emerged as a platform to study thermodynamics at the quantum scale~\cite{Campaioli2024,Alicki2013,Binder2015,Gemme2022,Shi2022}. A central focus in this field is how quantum resources and correlations such as coherence and entanglement can be utilized to enhance features of the QB such as charging power~\cite{Campaioli2017,Ferraro2018,Crescente2020,Ghosh2020,Ghosh2021,Quach2022,Ueki2022,Gemme2023} and maximum extractable work~\cite{Andolina2019,Barra2019,Kamin2020-2,Francica2020,Lai2024}. Various models have been used to study the behavior of QB charging, storage and work extraction~\cite{Le2018,Manzano2018,Rossini2019,Santos2019,Primoradian2019,Liu2019,Rossini2020,Quach2020,Monsel2020}.

Recently, a charging protocol employing a quantum interferometric approach called superposition of trajectories was explored~\cite{Lai2024}. Superposition of trajectories allows the system to travel along a coherent superposition of trajectories, such that the system experiences an effective higher order evolution~\cite{Chiribella2019, Foo2020}. It has been used to improve communication~\cite{Gisin2005,Kristjansson2020,Rubino2021,Ku2023}, thermodynamic performance~\cite{Ban2020,Chan2022}, metrology~\cite{Lee2023} and produce interesting open physics~\cite{Foo2021-2,Siltanen2021,Lin2022,Lin2023}. This approach was used to design protocols, which resulted in the QB being charged by multiple cavity chargers at the same time or interacting within a single cavity charger in multiple locations~\cite{Lai2024}. In both instances, the protocols display the \textit{quick charging effect}, whereby ergotropy can be generated immediately after charging begins. The effect originates from the coherent control of space-time trajectories, which utilizes coherence as a resource~\cite{Lai2024}. Similar usage of coherence as a resource in quantum thermodynamics has been explored as well~\cite{Aberg2014,Lostaglio2015,Lostaglio2015_2}.

In this work, we explore utilizing another quantum interferometric technique known as indefinite causal order (ICO)~\cite{Rubino2017,Goswami2018,Ebler2018,Zhao2020,Loizeau2020,Chiribella2021}. By employing a quantum switch, the causality of two or more events (such as interactions with baths or passage through quantum channels) is placed in coherent superposition. For the system experiencing these events, \textit{the exact sequence which the events happened cannot be definitely defined}, hence breaking the causality and earning the technique its name. Mathematically, ICO allows the system to effectively evolve through a higher order quantum map, which has shown advantages in information processing and thermodynamic tasks~\cite{Chiribella2013,Oreshkov2012,Araujo2014,Felce2020,Simonov2022,Francica2022-2}.

In this work, we construct a charging protocol that utilizes cyclic ICO~\cite{Procopio2019,Chiribella2021-2, Wang2025}, where the quantum switch coherently superposes all interaction sequences that belong to the cyclic group $\mathcal{Z}$. This brings about major differences compared with previous charging protocols utilizing ICO, as previous works focuses on utilizing superposition of two charging sequences for a finite number of chargers~\cite{Zhu2023,Li2025}. Here, the number of superposed charging sequences is equal to the number of chargers used.

We evaluate the performance of our cyclic ICO charging protocol by observing the difference in charging efficiency when compared to a definite causal order (DCO) charging protocol. Charging efficiency looks at the percentage of work that can be extracted from the QB after charging which is the ratio of the ergotropy to the energy stored~\cite{Biswas2022,MLSong2024}. The DCO charging protocol can be implemented with the same mathematical formulation, except the quantum switch is initialized in a pure and incoherent state such that the QB experiences a single charging sequence. Our cyclic ICO charging protocol exhibits charging efficiency bursts, \textit{whereby the percentage of extractable work increases dramatically for a short duration}. 

In addition to theoretical analysis and numerical simulations, we provide data from demonstrations on cloud quantum computing platforms to reinforce our findings. Explicitly, we provide a quantum circuit of our cyclic ICO charging protocol in the two-charger scenario tailored to gate-based quantum processors. We implement proof-of-concept demonstrations on IBM, IonQ, and Quantinuum quantum devices, demonstrating the feasibility of realizing cyclic ICO-enhanced QB charging in current noisy intermediate-scale quantum hardware~\cite{Preskill2018,Bharti2022} and observing the charging efficiency bursts. 

The rest of the paper is organized as follows. In Sec.~\ref{sec:ico}, we provide mathematical details and simulations of our cyclic ICO charging protocol. In Sec.~\ref{sec:circuit}, we consider the circuit implementation of our cyclic ICO charging protocol when utilizing two chargers and present the results using quantum processors of IBMQ, IonQ and Quantinuum. Finally, we draw our conclusions in Sec.~\ref{sec:summary}.

\section{Charging a qubit battery with cyclic indefinite causal order}\label{sec:ico}

\subsection{Cyclic indefinite causal order charging protocol}

\begin{figure}[!htbp]
\includegraphics[width=1\columnwidth]{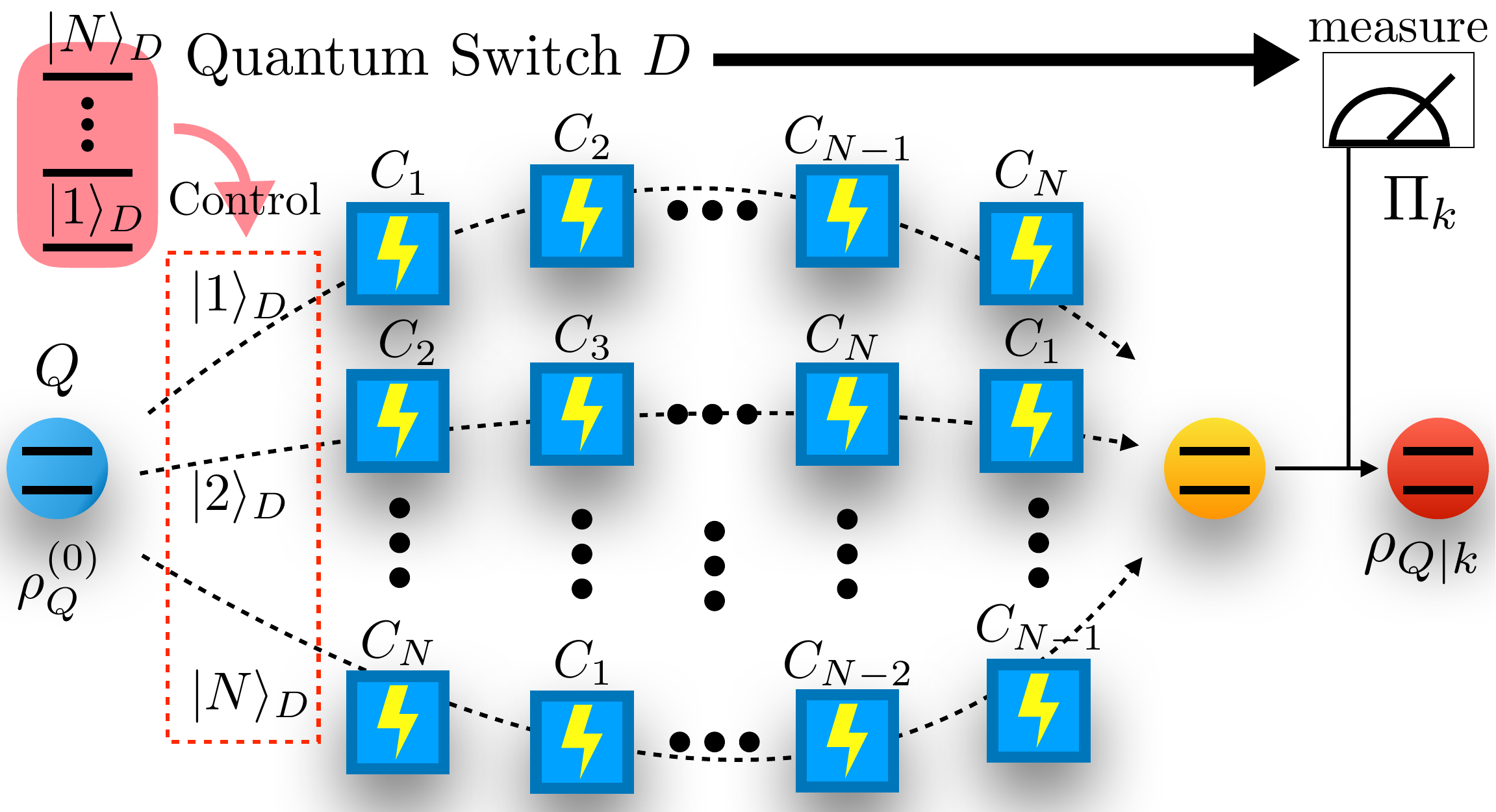}
\caption{Illustration of the cyclic indefinite causal order charging protocol for a qubit battery. The quantum battery $Q$ charges with the qubit chargers ($C_1\sim C_N$) in sequences that are elements of the cyclic group $\mathcal{Z}_N$. The charging sequence is controlled by the state $\ket{j}_D$ of the quantum switch $D$. With the quantum switch initialized in the superposition state $\sum_{m=1}^N\ket{m}_D/\sqrt{N}$, the quantum battery experiences a coherent superposition of all charging sequences which results in indefinite causal order. Therefore, the evolved state of the quantum switch $D$, quantum battery $Q$, and chargers $\{C_1,\cdots,C_N\}$ are $U_{\text{tot}}\rho_{\text{in}}U_{\text{tot}}^\dag$, where $\rho_{\text{in}}$ is the total initial state. Finally, measurement is performed on the quantum switch $D$ using projectors $\Pi_k$ to obtain the conditional quantum battery states $\rho_{Q|k}$, which is used for work extraction.}
\label{fig:protocol}
\end{figure}

In this section, we formulate our cyclic ICO charging protocol for a QB as shown in Fig.~\ref{fig:protocol}. The components utilized are: (i) A two-level QB $Q$. (ii) $N$ qubit chargers $\{C_1,\cdots,C_N\}$. (iii) A quantum switch $D$.

Our charging protocol is described by the unitary evolution of all parties $U_{\text{tot}}$.
\begin{equation}\label{eq:switch-unitary}
    U_{\text{tot}}=\sum_{j=1}^N\ket{j}\bra{j}_D\otimes V_j(\prod_{l=N}^1U_l),
\end{equation}
where $U_l$ represent the unitary interaction between the QB and charger $C_l$. Here, $V_j$ is an operation which permutes the unitaries $\prod_{l=N}^1U_l$ such that the charging sequence is the $j$th element of the cyclic group $\mathcal{Z}_N$. When the switch is in the $\ket{j}_D$ state, $Q$ interacts with the chargers via sequence $V_j(\prod_{l=N}^1U_l)$. To perform ICO, we initialize $D$ in a superposition state $\sum_{m=1}^N\ket{m}_D/\sqrt{N}$. We initialize the QB state as $\rho_Q^{(0)}$ and the chargers in states $\rho_{C_l}^{(0)}$ . The joint input is therefore
\begin{equation}\label{eq:input-state}
    \rho_{\text{in}} = \frac{1}{N}\sum_{m,n=1}^N\ket{m}\bra{n}_D\otimes\rho_Q^{(0)}\bigotimes_{l=1}^N\rho_{C_l}^{(0)}.
\end{equation}
After unitary evolution $U_{\text{tot}}$ (charging), we obtain $U_{\text{tot}}\rho_{\text{in}}U_{\text{tot}}^\dag$. Next, we measure the quantum switch $D$ with projectors $\{\Pi_k\}$. Conditioned on the outcome $k$, the (unnormalized) post-measurement state of the QB and chargers is
\begin{equation}\label{eq:postmeas-unnorm}
    \sigma_{Q \textbf{C}|k} = \text{Tr}_{D}\left[\Pi_k\,U_{\text{tot}}\,\rho_{\text{in}}\,U_{\text{tot}}^{\dagger}\,\Pi_k\right],
\end{equation}
with conditional probability $p_k=\text{Tr}\big[\sigma_{Q \textbf{C}|k}\big]$. 
The conditional QB state is
\begin{equation}\label{eq:cond-battery}
    \rho_{Q|k}=\text{Tr}_{\textbf{C}}[\sigma_{Q\textbf{C}|k}]/p_k,
\end{equation}
where $\text{Tr}_{\textbf{C}}$ traces out all the chargers. If the input of $D$ is $\ket{j}_D$ instead of a superposition state, this becomes a DCO charging protocol resulting in the QB state $\overline{\rho}_Q$. In the following, we use $\overline{\rho}_Q$ for comparison to the QB states $\rho_{Q|k}$ to observe charging efficiency burst from our cyclic ICO charging protocol. 
To compare their charging efficiency, we calculate their maximum extractable work known as ergotropy $W$~\cite{Alicki2013} and their energy stored within the QB after charging which we refer to as stored energy $E$. Charging efficiency is the ratio of the ergotropy to the stored energy $P=W/E$, which measures how efficient the charging protocol is at generating work that can be utilized. 

Let $H_Q$ denote the bare Hamiltonian of the QB. Thus, the ergotropy of the QB states $\rho_{Q|k}$ after our cyclic ICO charging protocol $W_{\text{ICO}}$ is
\begin{equation}
\begin{aligned}
    &W_{\text{ICO}}=\sum_k p_k W_k, \\ 
    &W_k=\mathrm{Tr}\left[\rho_{Q|k}H_Q\right]-\mathrm{Tr}\left[\varphi_{Q|k}H_Q\right],
\end{aligned}
\end{equation}
where $\varphi_{Q|k}$ is the passive state of $\rho_{Q|k}$ with respect to $H_Q$. This form of ergotropy is sometimes referred to as daemonic ergotropy or average ergotropy. 
The stored energy for $\rho_{Q|k}$ is 
\begin{equation}
\begin{aligned}
    &E_{\text{ICO}}=\sum_k p_k\mathrm{Tr}\left[\rho_{Q|k}H_Q\right] - \mathrm{Tr}\left[\rho_Q^{(0)}H_Q\right] \\
    &=\mathrm{Tr}\left[\sum_k p_k \rho_{Q|k}H_Q\right] - \mathrm{Tr}\left[\rho_Q^{(0)}H_Q\right].
\end{aligned}
\end{equation}
For a DCO charging protocol, the stored energy and ergotropy is 
\begin{equation}
\begin{aligned}
    &E_{\text{DCO}}=\mathrm{Tr}\left[\overline{\rho}_QH_Q\right] - \mathrm{Tr}\left[\rho_Q^{(0)}H_Q\right], \\
    &W_{\text{DCO}}=\mathrm{Tr}\left[\overline{\rho}_QH_Q\right]-\mathrm{Tr}\left[\overline{\varphi}_{Q}H_Q\right],
\end{aligned}
\end{equation}
where $\overline{\varphi}_{Q}$ is the passive state of $\overline{\rho}_Q$. The charging efficiency for our cyclic ICO charging protocol $P_\text{ICO}$ and DCO charging protocol $P_\text{DCO}$ is
\begin{equation}
\begin{aligned}
    &P_\text{ICO}=W_\text{ICO}/E_\text{ICO}, \\
    &P_\text{DCO}=W_\text{DCO}/E_\text{DCO}.
\end{aligned}
\end{equation}

% \medskip
% \noindent\textbf{Hamiltonian and initial state}\;
\subsection{Qubit charging with XY interaction}

To observe the difference in ergotropy when using ICO, we define the QB-charger Hamiltonian $H_l$ to obtain the unitary interactions $U_l=e^{-iH_l t/\hbar}$.
\begin{equation}\label{eq:hamiltonian}
    H_l=\frac{\hbar\omega}{2}(\hat{\sigma}_z^l+\openone^l)+\frac{\hbar\omega}{2}\hat{\sigma}_z^Q
    +\frac{\hbar\omega\lambda}{2}\left(\hat{\sigma}_x^Q\hat{\sigma}_x^l+\hat{\sigma}_y^Q\hat{\sigma}_y^l\right),
\end{equation}
where 
\begin{equation}
\begin{aligned}
&\sigma_z^{Q(l)}=\ket{e}\bra{e}_{Q(C_l)}-\ket{g}\bra{g}_{Q(C_l)}, \\
&\sigma_x^{Q(l)}=\ket{e}\bra{g}_{Q(C_l)}+\ket{g}\bra{e}_{Q(C_l)}, \\
&\sigma_y^{Q(l)}=-i\ket{e}\bra{g}_{Q(C_l)}+i\ket{g}\bra{e}_{Q(C_l)},
\end{aligned}
\end{equation}
with $\ket{e}$ being the excited state and $\ket{g}$ being the ground state. Here, $\omega$ is the on resonant frequency of $Q$ and $C_l$, $\lambda$ is the coupling strength between $Q$ and $C_l$. 

We can obtain the unitary evolution $U_l(t_l)$ between the QB and charger $l$ for some time $t_l$
\begin{equation}\label{eq:battery charger unitary}
\begin{aligned}
    &U_l(t_l)=e^{\frac{-3i\omega t_l}{2}}\ket{e}\bra{e}_Q\ket{e}\bra{e}_{C_l} \\
    &+e^{\frac{-i\omega t_l}{2}}\cos\omega\lambda t\ket{e}\bra{e}_Q\ket{g}\bra{g}_{C_l} \\
    &+e^{\frac{-i\omega t_l}{2}}(-i\sin\omega\lambda t_l)\ket{g}\bra{e}_Q\ket{e}\bra{g}_{C_l} \\
    &+e^{\frac{-i\omega t_l}{2}}(-i\sin\omega\lambda t_l)\ket{e}\bra{g}_Q\ket{g}\bra{e}_{C_l} \\
    &+e^{\frac{-i\omega t_l}{2}}\cos\omega\lambda t_l\ket{g}\bra{g}_Q\ket{e}\bra{e}_{C_l} \\ 
    &+e^{\frac{i\omega t_l}{2}}\ket{g}\bra{g}_Q\ket{g}\bra{g}_{C_l}.
\end{aligned}
\end{equation}
This allows us to obtain $U_{\text{tot}}(t)$, where $t=\sum_{l=1}^Nt_l$. By choosing the initial states to be $\rho_Q^{(0)}=\ket{g}\bra{g}_Q, \rho_{C_l}^{(0)}=\ket{e}\bra{e}_{C_l}$ and setting $t_l=t/N$, we obtain  the evolved state of the QB and chargers along the $j$th charging sequence $\rho_{Q\textbf{C}}^j(t)$ expressed using the basis $\ket{g,h_0}_{Q,\textbf{C}}=\ket{g}_Q\bigotimes_{l=1}^N\ket{e}_{C_l}$ and $\ket{e,h_l}_{Q,\textbf{C}}=\ket{e}_Q\hat{a}_l\bigotimes_{l=1}^N\ket{e}_{C_l}$ with time-dependent coefficients $\{\alpha_0(t),\alpha_1(t)\cdots,\alpha_N(t)\}$. The explicit form of $\rho_{Q\textbf{C}}^j(t)$ and its derivation is shown in Appendix~\ref{app:mathematical derivation}.
\begin{figure}[!htbp]
\includegraphics[width=1\columnwidth]{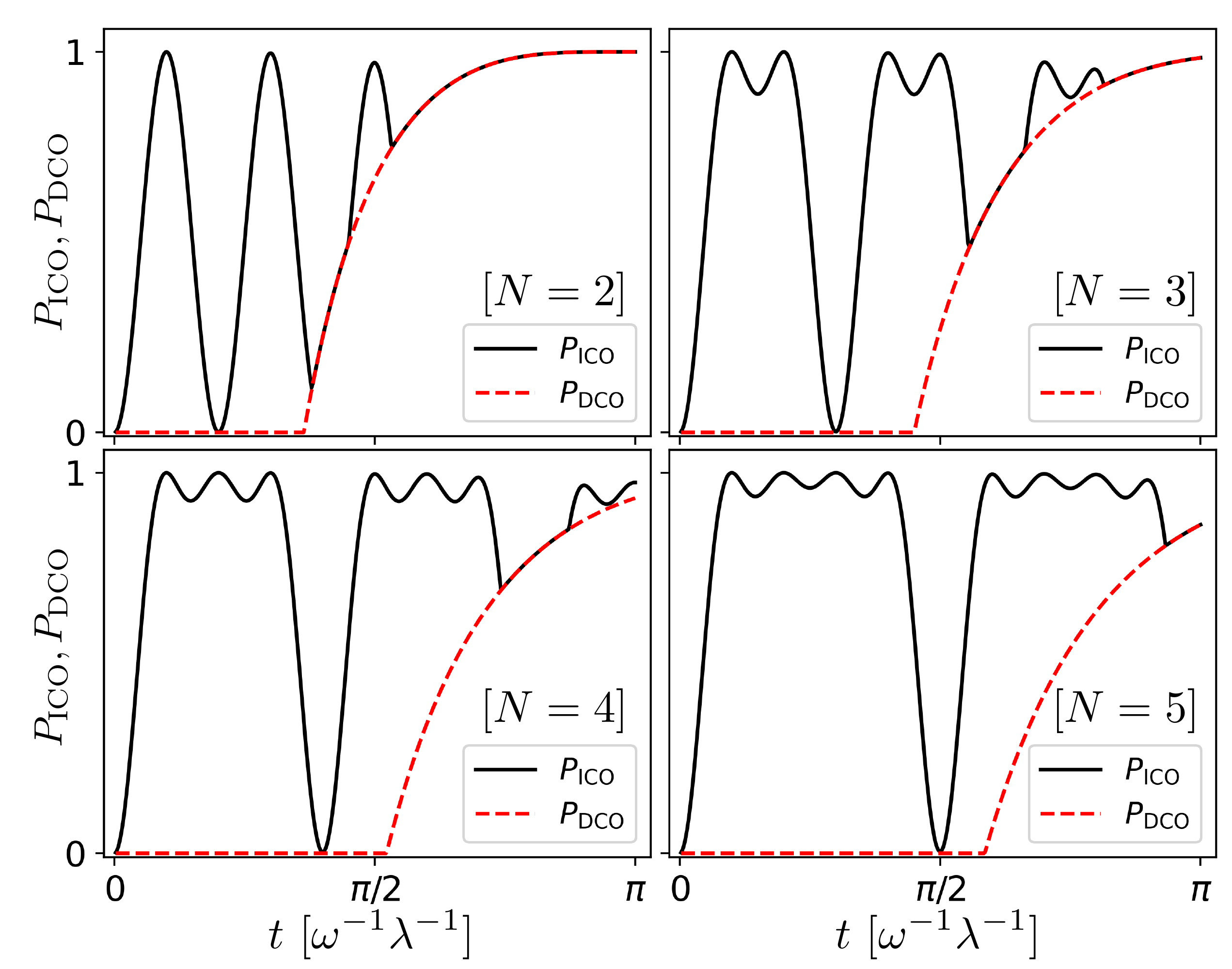}
\caption{Performance of qubit battery after charging time $t$. The black solid curves plot the charging efficiency using our cyclic indefinite charging protocol $P_{\text{ICO}}$, while the red dashed curves represent the charging efficiency using a definite charging protocol $P_\text{DCO}$. The results for using $N=2,3,4,5$ chargers in our protocols are shown.}
\label{fig:ico analytic}
\end{figure}
Next, for the projectors $\Pi_k$, we choose 
\begin{equation}
    \Pi_{k=1}=\frac{1}{N}\sum_{m,n=1}^N\ket{m}\bra{n}_D.
\end{equation}
The rest of the projectors can be chosen randomly as long as they are orthonormal to $\Pi_{k=1}$ and each other and satisfies completeness. This is due to the post-measurement QB state being $\ket{e}_Q$ when $\Pi_{k\neq1}$ is used, which we prove in Appendix~\ref{app:mathematical derivation}. The unnormalized post-measurement QB states conditioned on outcome $k$ are:
\begin{equation}\label{eq:postmeasurement}
\begin{aligned}
    &p_{k=1}(t)\rho_{Q|k=1}(t)=|\alpha_0(t)|^2\ket{g}\bra{g}_Q \\
    &+\left(\frac{C_{k=1}^N}{N}+\sum_{u=1}\frac{|\alpha_u(t)|^2}{N}\right)\ket{e}\bra{e}_Q, \\
    &\sum_{k\neq1}^Np_{k\neq1}(t)\rho_{Q|k\neq1}(t)=[1-p_{k=1}(t)]\ket{e}\bra{e}_Q,
\end{aligned}
\end{equation}
where, 
\begin{equation}
\begin{aligned}
    &C_{k=1}^N= \\
    &\sum_{u=1}^{N-1}\frac{2(N-u)}{N} \Re[\sum_{v=1}^N\alpha_v(t)\alpha^*_{((v\bmod N)+u)}(t)].
\end{aligned}
\end{equation}
and $\rho_{k\neq1}(t)=\ket{e}\bra{e}_Q$.
\begin{figure*}[!htbp]
\includegraphics[width=2\columnwidth]{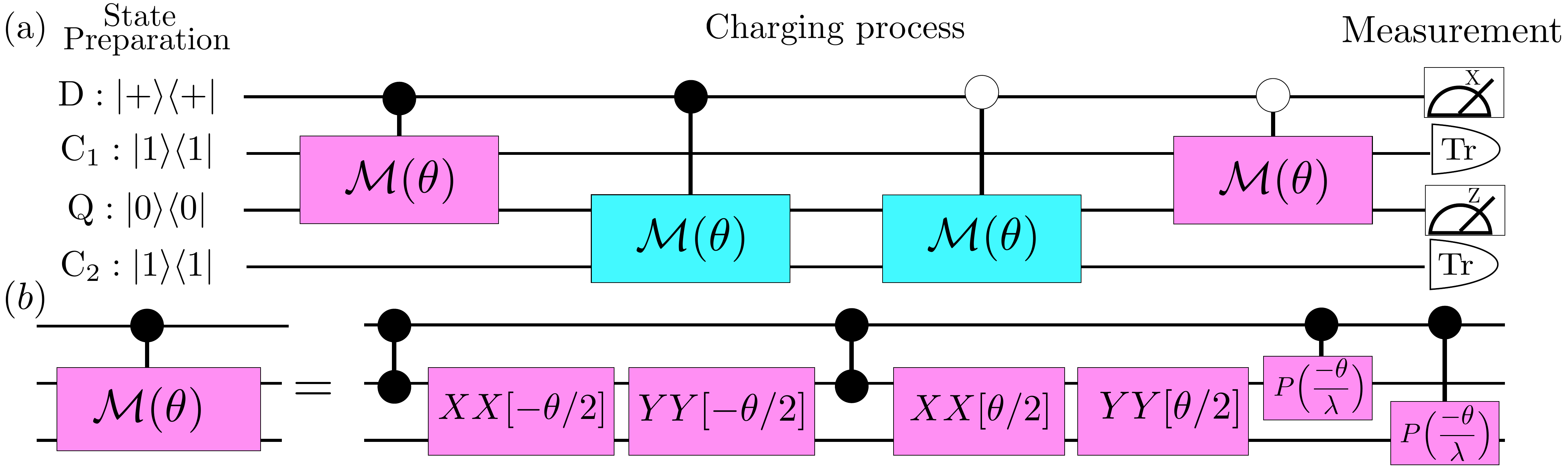}
\caption{(a) Quantum circuit for the two-charger cyclic indefinite charging protocol. Here, D,Q,C1,C2 represent the quantum switch, battery qubit, first charger qubit, and second charger qubit, respectively. “Tr” is trace out. (b) Decomposition of the controlled unitary $\mathcal{M}(\theta)$ in (a).}
\label{fig:ico circuit}
\end{figure*}
The stored energy $E_\text{ICO}$ is calculated to be 
\begin{equation}
    E_\text{ICO}=\hbar\omega\left[1-\left(\cos\frac{\omega\lambda t}{N}\right)^{2N}\right].
\end{equation}
Depending on whether $\rho_{Q|k=1}$ is a passive state or not, the ergotropy $W_{\text{ICO}}$ is
\begin{equation}
\begin{aligned}
    &W_{\text{ICO}}= \\
    &\begin{cases}
        \hbar\omega\!\left[\frac{N-1}{N}\!\sum_{j=1}^N|\alpha_j(t)|^2-\frac{C_{k=1}^N(t)}{N}\right], ~ \text{if passive}; \\
        \hbar\omega\!\left[1-2(\cos\frac{\omega\lambda t}{N})^{2N}\right],~\text{if not passive}.
    \end{cases} 
\end{aligned}
\end{equation}

For $\overline{\rho}_Q$, its stored energy is $E_\text{DCO}=E_\text{ICO}$, whereas its ergotropy is
\begin{equation}
\begin{aligned}
    &W_{\text{DCO}}=
    \begin{cases}
        0,~~
        \text{if}~\overline{\rho}_Q~\text{is passive}; \\
        \hbar\omega\!\left[1-2(\cos\frac{\omega\lambda t}{N})^{2N}\right],~\text{if not}.
    \end{cases}
\end{aligned}
\end{equation}
The mathematical details of the above calculations are presented in Appendix~\ref{app:mathematical derivation}.

We plot the charging efficiency utilizing cyclic ICO $P_\text{ICO}$ and the charging efficiency utilizing DCO $P_\text{DCO}$ with respect to the total charging time $t$ in Fig.~\ref{fig:ico analytic}. The black solid curves represent the charging efficiency using our cyclic ICO charging protocol $P_{\text{ICO}}$. The red dashed curves represent the charging efficiency using our DCO charging protocol $P_{\text{DCO}}$. The results for $N=2,3,4,5$ are shown. Compared with the charging efficiency using DCO, the charging efficiency using our cyclic ICO charging protocol exhibits bursts of improvement. Notably, $P_\text{DCO}$ remains at zero for some time, whilst $P_\text{ICO}$ exhibit high (though not perfect) efficiency bursts for that duration. We also observe that the duration of charging efficiency burst for our cyclic ICO charging protocol increases with $N$.

\section{Implementation on quantum devices}\label{sec:circuit}
%Two-qubit gate number: Rigetti 70; IQM 52; IonQ 40; IBMQ 56.
%Rigetti Ankaa-2: 35 circuit number (points.)1000 shots; IQM Garnet: 25 circuit number (points.)1000 shots; IBMQ Marrakesh 200 points (originally), 1000 shots; quantinuum 35 ponts, 150 shots; IonQ Aria-2: 35 points, 1000 shots. Those are data details that I ran before. (I saw you wrote all the machines from different companies in the manuscript so I provided all of them.)

In this section, we provide a circuit model of the proposed cyclic ICO charging protocol for the case of $N=2$. We then perform proof-of-concept demonstrations on quantum processors provided by IBMQ, IonQ, and Quantinuum. The backends used and their relevant properties are detailed in Appendix~\ref{app:backend}.

The quantum circuit of the $N=2$ cyclic ICO charging protocol is shown in Fig.~\ref{fig:ico circuit}(a). The circuit consists of four qubits: quantum switch $D$, the QB $Q$, and two charging qubits $C_1,C_2$. The circuit is divided into three parts: state preparation, charging process and measurement on the quantum switch and QB. During state preparation, the qubits are initialized as follows with single qubit gates:
\begin{equation}
\begin{aligned}
    &\text{quantum switch}~D:\ket{+}_D, \\
    &\text{quantum battery}~Q:\ket{0}_Q, \\
    &\text{qubit charger 1}~C_1:\ket{1}_{C_1}, \\
    &\text{qubit charger 2}~C_2:\ket{1}_{C_2},
\end{aligned}
\end{equation}
where we take $\ket{g}=\ket{0}, \ket{e}=\ket{1}$.  The charging process involves the utilization of four controlled-unitary gates to simulate the ICO of charging sequences through XY interactions. In Fig.~\ref{fig:ico circuit}(b), we present the decomposition of the controlled unitaries into controlled-z gates (CZ), Ising coupling gates (XX and YY) and controlled-phase gates (CP), defined as
\begin{equation}
\begin{aligned}
    &CZ=\ket{0}\bra{0}\otimes\id + \ket{1}\bra{1}\otimes\hat{\sigma}_z, \\
    &XX(\theta)=\cos(\theta/2)\id\otimes\id-i\sin(\theta/2)\hat{\sigma}_x\otimes\hat{\sigma}_x, \\
    &YY(\theta)=\cos(\theta/2)\id\otimes\id-i\sin(\theta/2)\hat{\sigma}_y\otimes\hat{\sigma}_y, \\
    &CP(\theta)=\ket{0}\bra{0}\otimes\id + \ket{1}\bra{1}\otimes \mathcal{P},
\end{aligned}
\end{equation}
where we write $\mathcal{P}=\ket{0}\bra{0}+e^{i\theta}\ket{1}\bra{1}$. Here, we map the total charging time $t$ into the angle $\theta$ using
\begin{equation}
    \theta(t)=\omega\lambda t/2.
%學長我們這裡的角度跟時間的轉換應該是用 \theta(t)=\omega\lambda t 就是說時間沒有除以二 原因是因為充電時間跟之前學長的PRR是不一樣長的(X軸的時間變兩倍這樣)
\end{equation}
Finally, we measure the quantum switch $D$ in the x-direction, which corresponds to measuring in the $\{\ket{+}=(\ket{0}
+\ket{1})/\sqrt{2},\ket{-}=(\ket{0}-\ket{1})/\sqrt{2}\}$ basis. Furthermore, as indicated from Eq.~(\ref{eq:postmeasurement}), the post-measurement states are diagonalized under the energy eigenstate basis. Thus, measurement results of $Q$ in the z-basis allow us to derive $E$ and $W_{\text{ICO}}$.

\begin{figure}[!htbp]\label{fig:exp}
\includegraphics[width=1\columnwidth]{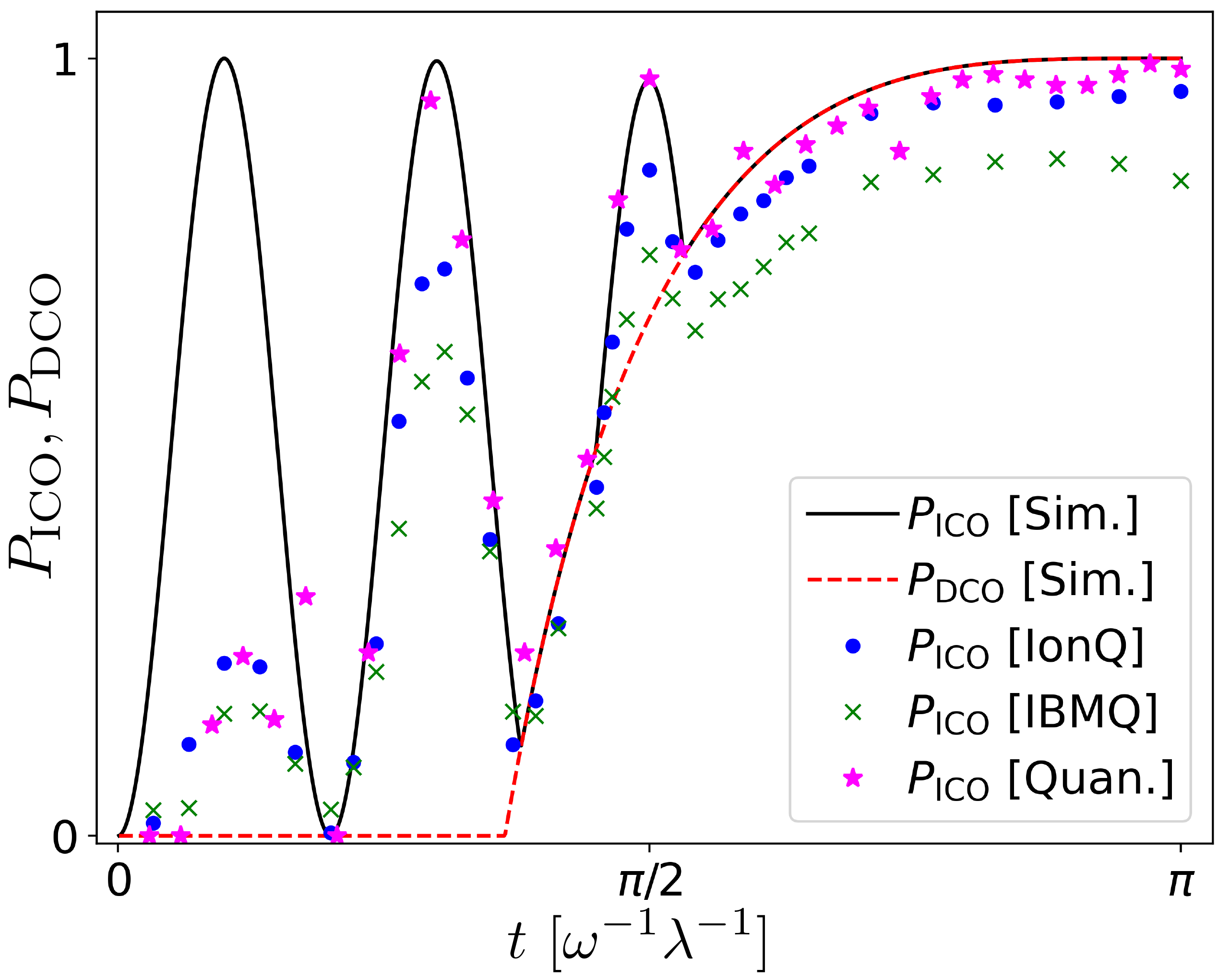}
\caption{Charging efficiency of the cyclic indefinite causal order charging protocol with two chargers on quantum devices with respect to total charging time t. The black solid curve represent the charging efficiency of our cyclic indefinite causal order charging protocol $P_\text{ICO}$ predicted by numerical simulations. The red dashed curve represent the charging efficiency of our definite causal order charging protocol $P_\text{DCO}$ predicted by numerical simulations. The blue circles represent results performed on IonQ. The green x represent results performed on IBMQ. The magenta stars represent results performed on Quantinuum. Repetitions for IonQ, IBMQ, Quantinuum are 240000, 20000, 150, respectively. Backend properties are shown in Appendix.~\ref{app:backend}.}
\end{figure}

We present the results of our demonstrations in Fig.~\ref{fig:exp}. For IonQ, IBMQ, Quantinuum results, each data point represents 240000, 20000, 150 repetitions, respectively. The theoretical predictions for the charging efficiency using cyclic ICO and DCO are shown as the black solid curve and the red dashed curve, respectively. We observe that the blue circle, green x and magenta star markers which represent the charging efficiency of our cyclic ICO charging protocol calculated with IonQ, IBMQ, and Quantinuum processors all demonstrate charging efficiency bursts to different degrees. Particularly, they all demonstrate the possibility of extracting work during times when $P_\text{DCO}$ is zero. Thus, quantum processors can demonstrate charging efficiency burst using our cyclic ICO charging protocol. 

We notice that the error is greater near the beginning of the cyclic indefinite charging protocol, particularly the first burst. We find that around the first burst, the results greatly overestimate the stored energy $E_\text{ICO}$, leading to a significant drop in charging efficiency $P_\text{ICO}$. This can be attributed to depolarizing error as a result of accumulating gate error, which is modeled by incoherently mixing the ideal system’s state with the maximally mixed state~\cite{Magesan2011}. We note that the ideal QB state at the beginning and end is $\ket{g}$ and $\ket{e}$, respectively. This makes them more susceptible to depolarizing error. In contrast, the ideal states in the middle of charging are naturally mixed states, making them more resilient to depolarizing error~\cite{Lai2024}. 

\section{Summary and Outlook}\label{sec:summary}
We propose a charging protocol for quantum batteries that utilizes cyclic indefinite causal order which exhibits charging efficiency burst. We provide analytical calculations and numerical simulations for a qubit battery being charged by $N$ qubit chargers in an indefinite sequence. We find bursts of charging efficiency when compared with the results of using a definite charging protocol. The duration of these bursts increase with the number of qubit chargers used. We further provide a circuit model for the two-charger scenario of our cyclic indefinite causal order charging protocol and perform demonstrations on IonQ, IBMQ, Quantinuum quantum processors. The results validate our theoretical predictions. 

Possible future directions to extend our indefinite causal order charging protocol include exchanging the charging media from qubits to cavities or waveguides~\cite{Song2024,Tirone2025,Guo2025,Salvia2023,Rinaldi2025}. It would also be interesting to explore whether similar charging efficiency bursts could exist when charging a quantum battery in the presence of non-Markovian memory effects~\cite{Kamin2020-2,Morrone2023}.

\section*{Acknowledgments}
We acknowledge the NTU-IBM Q Hub, IBM quantum experience, and Cloud Computing Center for Quantum Science $\&$ Technology at NCKU for providing us a platform to implement the demonstrations. This work is supported by the National Center for Theoretical Sciences and National Science and Technology Council, Taiwan, Grants No. NSTC 114-2112-M-006-015-MY3.

\appendix
\section{Details of analytical derivation}\label{app:mathematical derivation}
In this section, we provide the mathematical details towards deriving the results in Sec.~\ref{sec:ico}~B.

We start by working with Eq.~(\ref{eq:battery charger unitary}), and utilizing the fact that the initial states of the QB and chargers are pure states: $\ket{g}_Q$ and $\ket{e}_{C_l}$, respectively, as well as $t_l=t/N$ to obtain the time-evolved wavefunctions of the QB and chargers along the $j$th charging sequence $\ket{\psi_j(t)}_{Q,\textbf{C}}$ such that $\rho_{Q\textbf{C}}^j(t)=\ket{\psi_j(t)}\bra{\psi_j(t)}_{Q,\textbf{C}}$.

From the QB and charger initial states, their time-evolved wavefunction only spans the basis $\ket{g,h_0}_{Q,\textbf{C}}=\ket{g}_Q\bigotimes_{l=1}^N\ket{e}_{C_l}$ and $\ket{e,h_l}_{Q,\textbf{C}}=\ket{e}_Q\hat{a}_l\bigotimes_{l=1}^N\ket{e}_{C_l}$ for $l=1\sim N$. We notice that for $j=1$:
\begin{equation}
\begin{aligned}
    &\ket{\psi_{j=1}(t)}_{Q,\textbf{C}}=V_{j=1}(\prod_{l=N}^1U_l(t/N))\ket{g,h_0}_{Q,\textbf{C}} \\
    &=U_NU_{N-1}\cdots U_1\ket{g,h_0}_{Q,\textbf{C}} \\
    &=(e^{\frac{-i\omega t}{2N}}\cos\frac{\omega\lambda t}{N})^{N}\ket{g,h_0}_{Q,\textbf{C}}\\
    &+\sum_{j=1}^N\left[\Big(e^{\frac{-3i\omega t}{2N}}\Big)^{N-j}e^{\frac{-i\omega t}{2N}}\Big(-i\sin\frac{\omega\lambda t}{N}\Big)\times\right. \\
    &\left.\Big(e^{\frac{-i\omega t}{2N}}\cos\frac{\omega\lambda t}{N}\Big)^{j-1}\right]\ket{e,h_j}_{Q,\textbf{C}} \\
    &=\alpha_0(t)\ket{g,h_0}_{Q,\textbf{C}}+\sum_{j=1}^N\alpha_j(t)\ket{e,h_j}_{Q,\textbf{C}} \\
    &=\alpha_0(t)\ket{g,h_0}_{Q,\textbf{C}}+\text{diag}[\alpha_1(t),\cdots,\alpha_N(t)] \\
    &\times[\ket{e,h_1}_{Q,\textbf{C}},\cdots,\ket{e,h_N}_{Q,\textbf{C}}]^{\text{T}},
\end{aligned}
\end{equation}
where we choose the charging sequence for $j=1$ to be $1,2,3\cdots,N$ (trivial setting as we can always relabel the chargers). For simplicity, we replace the analytical time-dependent coefficients with $\alpha_{0\sim N}(t)$. We also write $\sum_{j=1}^N\alpha_j(t)\ket{e,h_j}_{Q,\textbf{C}}$ as multiplication between a diagonal matrix with elements $\{\alpha_1(t),\cdots,\alpha_N(t)\}$ and a column matrix with elements $\{\ket{e,h_1}_{Q,\textbf{C}},\cdots,\ket{e,h_N}_{Q,\textbf{C}}$. To obtain $\ket{\psi_j(t)}_{Q,\textbf{C}}$, we simply rearrange the elements in the column matrix. Thus,
\begin{equation}
\begin{aligned}
    &\ket{\psi_j(t)}_{Q,\textbf{C}} \\
    &=\alpha_0(t)\ket{g,h_0}_{Q,\textbf{C}}+\text{diag}[\alpha_1(t),\cdots,\alpha_N(t)]\times  \\
    &\overline{V}_j([\ket{e,h_1}_{Q,\textbf{C}},\cdots,\ket{e,h_N}_{Q,\textbf{C}}])^{\text{T}},
\end{aligned}
\end{equation}
where $\overline{V}_j$ rearranges the elements in the row matrix $[\ket{e,h_1}_{Q,\textbf{C}},\cdots,\ket{e,h_N}_{Q,\textbf{C}}]$ according to the $j$th element in the cyclic group $\mathcal{Z}_N$. With the quantum switch $D$ initially in the superposed state $\sum_{m=1}^N\ket{m}_D/\sqrt{N}$, the evolved state of the switch-QB-chargers is
\begin{equation}
\begin{aligned}
    &\rho_{\text{out}}(t)=U_{\text{tot}}(t)\rho_{\text{in}}U_{\text{tot}}^\dag(t) \\
    &=\frac{1}{N}\sum_{j,f=1}^N\ket{j}\bra{f}_D\otimes\ket{\psi_j(t)}\bra{\psi_f(t)}_{Q,\textbf{C}}.
\end{aligned}
\end{equation}

Next, we need to define the projectors which are orthonormal and complete:
\begin{equation}
\begin{aligned}
    &\Pi_k=\ket{\xi_k}\bra{\xi_k}_D,~\sum_{k=1}^N\Pi_k=\openone, \\
    &~\braket{\xi_k|\xi_{k'}}=\delta_{k,k'}~\forall~k,k'.
\end{aligned}
\end{equation}
We also set 
\begin{equation}
    \Pi_{k=1}=\frac{1}{N}\sum_{m,n=1}^N|m\rangle\langle n|_D,
\end{equation}
and leave the other projectors unspecified as their exact form is irrelevant to the post-measurement QB state, which we derive as follows. The unnormalized post-measurement QB states conditioned on outcome $k$ are:
\begin{equation}\label{eq:n charger post-measure state}
\begin{aligned}
    &\sigma_{Q|k}(t)=\text{Tr}_{D,\textbf{C}}[\Pi_k\rho_{\text{out}}\Pi_k] \\
    &=\frac{1}{N}\sum_{j,f=1}\braket{\xi_k|j}\braket{f|\xi_k}\text{Tr}_{\textbf{C}}[\ket{\psi_j}\bra{\psi_f}_{Q,\textbf{C}}] \\
    &=\frac{1}{N}\sum_{j=1}|\braket{\xi_k|j}|^2\text{Tr}_{\textbf{C}}[\ket{\psi_j}\bra{\psi_j}_{Q,\textbf{C}}] \\
    &+\frac{1}{N}\sum_{j>f}\sum_{f=1}\left[\braket{\xi_k|j}\braket{f|\xi_k}\text{Tr}_{\textbf{C}}[\ket{\psi_j}\bra{\psi_f}_{Q,\textbf{C}}]\right. \\
    &\left.+\braket{\xi_k|f}\braket{j|\xi_k}\text{Tr}_{\textbf{C}}[\ket{\psi_f}\bra{\psi_j}_{Q,\textbf{C}}]\right].
\end{aligned}
\end{equation}
We find that 
\begin{equation}
\begin{aligned}
    &\text{Tr}_{\textbf{C}}[|\psi_j(t)\rangle\langle\psi_j(t)|_{Q,\textbf{C}}] \\
    &=|\alpha_0(t)|^2|g\rangle\langle g|_Q+\sum_{j=1}^N|\alpha_j(t)|^2|e\rangle\langle e|_Q, \\
    &\text{Tr}_{\textbf{C}}[|\psi_j(t)\rangle\langle\psi_f(t)|_{Q,\textbf{C}}] \\
    &=|\alpha_0(t)|^2|g\rangle\langle g|_Q+X_{jf}(t)|e\rangle\langle e|_Q,
\end{aligned}
\end{equation}
where 
\begin{equation}
\begin{aligned}
    &X_{jf}(t)=\overline{V}_j([\alpha_1(t),\cdots,\alpha_N(t)]) \\
    &\times\overline{V}_f([\alpha_1^*(t),\cdots,\alpha_N^*(t)])^{\text{T}}.
\end{aligned}
\end{equation} 
Inserting back to Eq.~(\ref{eq:n charger post-measure state}), we obtain
\begin{equation}
\begin{aligned}
    &\sigma_{Q|k}(t)= \\
    &\left(|\alpha_0|^2|g\rangle\langle g|_Q+\sum_{j=1}^N|\alpha_j|^2|e\rangle\langle e|_Q\right)\frac{1}{N}\sum_{j=1}^N|\langle\xi_k|j\rangle|^2 \\
    &+\frac{1}{N}\sum_{j>f}\sum_{f=1}\left[\braket{\xi_k|j}\braket{f|\xi_k}(|\alpha_0|^2\ket{g}\bra{g}_Q\right. \\
    &+X_{jf}\ket{e}\bra{e}_Q) \\
    &\left.+\langle\xi_k|f\rangle\langle j|\xi_k\rangle(|\alpha_0|^2|g\rangle\langle g|_Q+X_{jf}^*|e\rangle\langle e|_Q)\right].
\end{aligned}
\end{equation}

Next, we solve $\sigma_{Q|k}(t)$ for the cases of $k=1$ and $k\neq1$. Here, $\sum_{j=1}|\langle\xi_k|j\rangle|^2=1~\forall~k$ due to orthonormality of projectors. Because the $\ket{\xi_{k\neq1}}$ need to be orthogonal to $\ket{\xi_{k=1}}$, they can be expressed as
\begin{equation}
    |\xi_{k\neq1}\rangle_D=\sum_{j=1}^N\beta_{k,j}|j\rangle_D,  
\end{equation}
where
\begin{equation}
    \sum_{j=1}^N\beta_{k,j}=0~\forall~k\neq1.
\end{equation}
Due to these properties, we can derive
\begin{equation}
    \sum_{j>f}\sum_{f=1}\langle\xi_{k\neq1}|j\rangle\langle f|\xi_{k\neq1}\rangle+\langle\xi_{k\neq1}|f\rangle\langle j|\xi_{k\neq1}\rangle= -1,
\end{equation}
whereas by making use of $\langle\xi_{k=1}|j\rangle=1/\sqrt{N}$ results in
\begin{equation}
\begin{aligned}
    &\sum_{j>f}\sum_{f=1}\langle\xi_{k=1}|j\rangle\langle f|\xi_{k=1}\rangle+\langle\xi_{k=1}|f\rangle\langle j|\xi_{k=1}\rangle \\
    &=N-1.
\end{aligned}
\end{equation}
Thus,
\begin{equation}
\begin{aligned}
    &\sigma_{Q|k=1}(t) \\
    &=\frac{1}{N}\Big(|\alpha_0(t)|^2|g\rangle\langle g|_Q+\sum_{j=1}^N|\alpha_j(t)|^2|e\rangle\langle e|_Q\big) \\
    &+\frac{N-1}{N}|\alpha_0(t)|^2|g\rangle\langle g|_Q \\
    &+\frac{1}{N}\sum_{j>f}\sum_{f=1}\left[\braket{\xi_{k=1}|j}\braket{f|\xi_{k=1}} X_{jf}(t)\right. \\
    &\left.+\braket{\xi_{k=1}|f}\braket{j|\xi_{k=1}} X_{jf}^*(t)\right]|e\rangle\langle e|_Q \\
    &=|\alpha_0(t)|^2|g\rangle\langle g|_Q \\
    &+\left[\frac{1}{N}C_{k=1}^N(t)+\frac{1}{N}\sum_{j=1}^N|\alpha_j(t)|^2\right]|e\rangle\langle e|_Q.
\end{aligned}
\end{equation}
Here, 
\begin{equation}
\begin{aligned}
    &C_{k}^N(t)=  \sum_{j>f}\sum_{f=1}\Big[\braket{\xi_{k}|j}\braket{f|\xi_{k}} X_{jf}(t) \\
    &+\braket{\xi_{k}|f}\braket{j|\xi_{k}}X_{jf}^*(t)\Big].
\end{aligned}
\end{equation}
Thus, the normalized state that we receive after measuring with projector $\Pi_{k=1}$ is $\rho_{Q|k=1}(t)=\sigma_{Q|k=1}(t)/\text{Tr}[\sigma_{Q|k=1}(t)]$.

For $k\neq1$:
\begin{equation}
\begin{aligned}
    &\sigma_{Q|k\neq1}(t) \\
    &=\frac{1}{N}(|\alpha_0(t)|^2|g\rangle\langle g|_Q+\sum_{j=1}^N|\alpha_j(t)|^2|e\rangle\langle e|_Q) \\
    &-\frac{1}{N}|\alpha_0(t)|^2|g\rangle\langle g|_Q+\frac{C_{k}^N(t)}{N}|e\rangle\langle e|_Q \\
    &=\Big(\frac{C_{k}^N(t)}{N}+\sum_{j=1}^N\frac{|\alpha_j(t)|^2}{N}\Big)|e\rangle\langle e|_Q.
\end{aligned}
\end{equation}
Note that we cannot easily derive a closed-form expression for  $\sigma_{Q|k\neq1}$. However, we can observe that it is proportional to the excited state $|e\rangle\langle e|_Q$. Thus, we need not specify the projectors $\Pi_{k\neq1}$ since they all return the excited state with different probability. The relevant property is their total probability which is $1-\text{Tr}[\sigma_{Q|k=1}(t)]$. Therefore,
\begin{equation}
    \sum_{k\neq1}^N\sigma_{k\neq1}(t)=(1-\text{Tr}[\sigma_{Q|k=1}(t)])|e\rangle\langle e|_Q,
\end{equation}
and $\rho_{Q|k\neq1}(t)=|e\rangle\langle e|_Q$.

With all the post-measurement QB states calculated, we can now obtain the stored energy $E_\text{ICO}$
\begin{equation}
\begin{aligned}
    &\hbar\omega E_\text{ICO} 
    =\text{Tr}\Big[\frac{\hat{\sigma}_z}{2}\Big(\sigma_{Q|k=1}(t) \\
    &+(1-\text{Tr}[\sigma_{Q|k=1}(t)])|e\rangle\langle e|_Q\Big)\Big] 
    - \text{Tr}\Big[\frac{\hat{\sigma}_z}{2}|g\rangle\langle g|_Q\Big] \\
    &=\text{Tr}\Big[\frac{\hat{\sigma}_z}{2}(|\alpha_0(t)|^2|g\rangle\langle g|_Q \\
    &+(1-|\alpha_0(t)|^2)|e\rangle\langle e|_Q)\Big] 
    -\text{Tr}\Big[\frac{\hat{\sigma}_z}{2}|g\rangle\langle g|_Q\Big] \\
    &=1-|\alpha_0(t)|^2=1-\left(\cos\frac{\omega\lambda t}{N}\right)^{2N}.
\end{aligned}
\end{equation}

For the ergotropy, we have two scenarios, one where $\sigma_{Q|k=1}$ is passive and one when it's not. If it's passive, then
\begin{equation}
\begin{aligned}
    &W_\text{ICO}=\hbar\omega(1-\text{Tr}[\sigma_{Q|k=1}(t)]) \\
    &=\hbar\omega\Big(\frac{N-1}{N}\sum_{j=1}^N|\alpha_j(t)|^2-\frac{C_{k=1}^N(t)}{N}\Big).
\end{aligned}
\end{equation}
If it's not, then
\begin{equation}
\begin{aligned}
    &W_\text{ICO}=\hbar\omega(1-2|\alpha_0(t)|^2) \\
    &=\hbar\omega\Big[1-2\left(\cos\frac{\omega\lambda t}{N}\right)^{2N}\Big].
\end{aligned}
\end{equation}

By once again using $\braket{\xi_{k=1}|j}=1/\sqrt{N}$, we can write a general compact form for $C_{k=1}^N(t)$.
\begin{equation}
\begin{aligned}
    &C_{k=1}^N(t) \\
    &=\frac{2}{N}\sum_{l=1}^{N-1}(N-l)\Re\Big[\sum_{j=1}^N\alpha_j\alpha_{((j\bmod N)+l)}^*\Big].
\end{aligned}
\end{equation}

If we utilize DCO, we obtain the mixed state $\bar{\rho}_Q=|\alpha_0|^2|g\rangle\langle g|+(1-|\alpha_0|^2)|e\rangle\langle e|$ which has energy $E_\text{DCO}=E_\text{ICO}$. Its ergotropy is
\begin{equation}
    W=
    \begin{cases}
        \hbar\omega(1-2|\alpha_0(t)|^2),&\text{if active} \\
        0,&\text{if passive}.
    \end{cases}
\end{equation}

\section{Backend properties of quantum processors used}\label{app:backend}
We utilized three quantum processors each from IBMQ, IonQ and Quantinuum to demonstrate the two-charger scenario of our cyclic indefinite causal order charging protocol in the main text. Here, we provide some details on the backend properties and connectivity used when running the demonstrations. 

\begin{table*}[!htbp]
  \centering
  \caption{\textit{ibm\_boston} calibration data (obtained Mar. 20, 2026). CZ and RZZ gates are non-directional, so the CZ and RZZ gate errors of qubit 123 are those listed on the rows of qubit 122, 124, 136.  }\label{table:ibm_boston}
  \begin{tabular}{c c c c c c c}
    \hline\hline
      Qubit label& Qubit index & \shortstack{\rule{0pt}{2.6ex} Relaxation time\\ $T_1$ ($\mu s$)} & \shortstack{Decoherence time\\ $T_2$ ($\mu s$)} & \shortstack{Readout\\ error} & \shortstack{CZ\\ error} & \shortstack{RZZ \\error}\\[0.4em] \hline
    $D$ & 122 & 194.54 & 209.32 & 5.85$\times10^{-3}$ & 1.29$\times10^{-3}$ & 1.53$\times10^{-3}$ \\[0.2em] 
    $Q$ & 123 & 300.89 & 253.39 & 5.12$\times10^{-3}$ &$\times$ & $\times$ \\[0.2em]
    $C_1$ & 124 & 284.84 & 443.48 & 7.56$\times10^{-3}$ & 7.75$\times10^{-4}$ & 8.44$\times10^{-4}$ \\[0.2em]
    $C_2$ & 136 & 315.11 & 338.86 & 4.63$\times10^{-3}$ & 1.07$\times10^{-3}$ & 1.04$\times10^{-3}$ \\[0.2em]
    
    \hline\hline
  \end{tabular}
\end{table*}

\subsection{IBMQ}
\begin{figure}[!htbp]
\includegraphics[width=1\columnwidth]{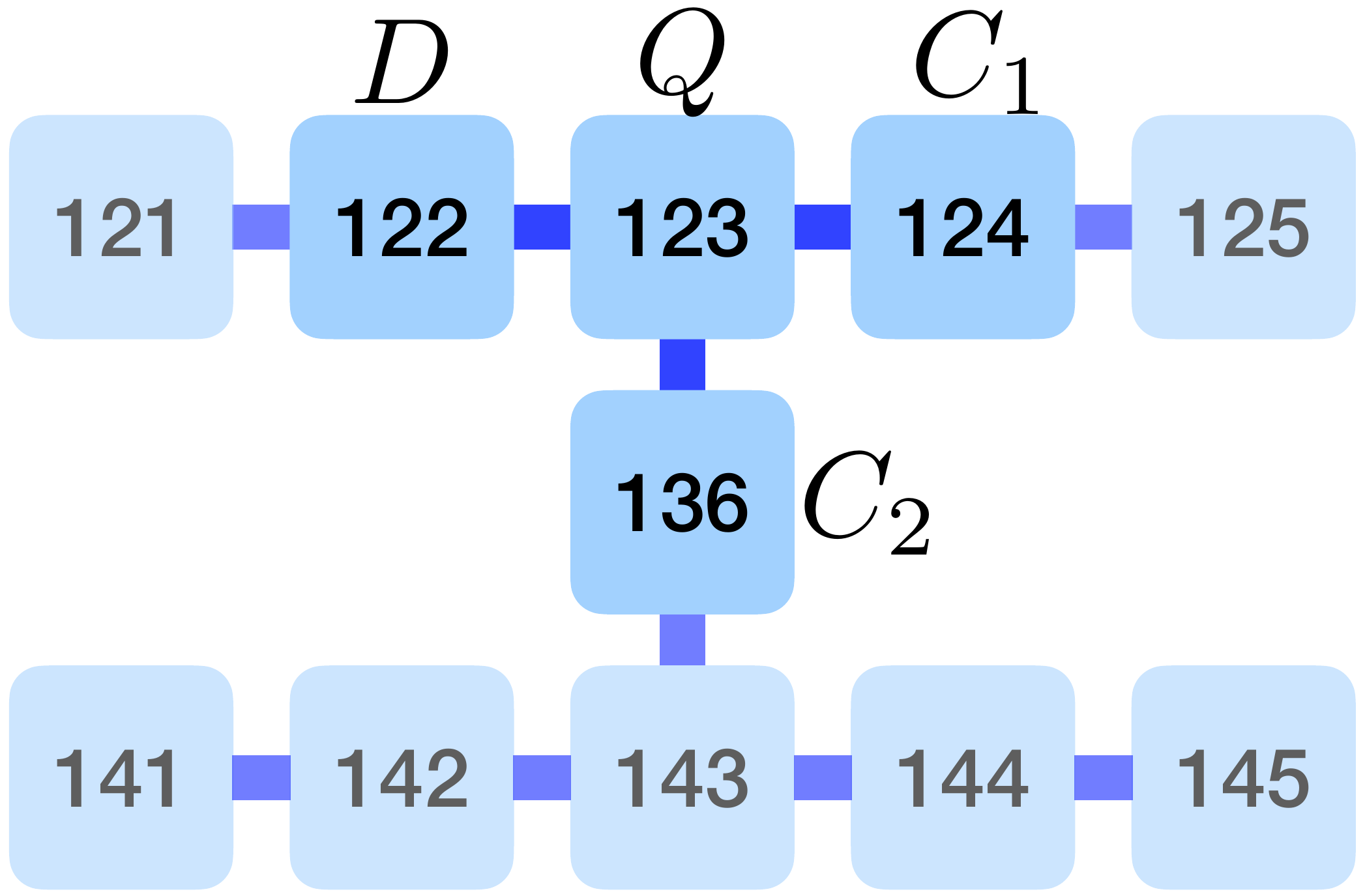}
\caption{The configuration of the qubits used on $ibm\_boston$ in our circuit. The control qubit $D$ used qubit 122, the battery qubit $Q$ used qubit 123, the first charger qubit $C_1$ used qubit 124 and the second charger qubit $C_2$ used qubit 136.}
\label{fig:ibmq configuration}
\end{figure}

For our demonstrations on IBMQ, we used the backend $\textit{ibm\_boston}$ which contains 156 superconducting qubits. We place the 4 qubits used in a T-shaped configuration as shown in Fig.~\ref{fig:ibmq configuration}. The calibration data of the qubits used is shown in Table~\ref{table:ibm_boston}.

\begin{table}[!htbp]
  \centering
  \caption{\textit{IonQ-Aria-1} calibration data (obtained Aug. 16, 2025).}\label{table:ionq aria 1}
  \begin{tabular}{c c}
    \hline\hline
      Relaxation time $T_1$ & 10$\sim$100 seconds \\[0.2em]
      Decoherence time $T_2$ & 1 second \\[0.2em]
      SPAM error & 3.9$\times10^{-3}$ \\[0.2em]
      Single-qubit gate error & 6$\times10^{-4}$ \\[0.2em]
      Two-qubit gate error & 6$\times10^{-3}$ \\[0.2em]
    
    \hline\hline
  \end{tabular}
\end{table}

\subsection{IonQ}
For our demonstrations on IonQ, we used the backend $\textit{IonQ-Aria-1}$ which contains 25 trapped ions. The all-to-all connectivity of the device allows us to pick any 4 qubits to perform our demonstration. The calibration data of the device is shown in Table~\ref{table:ionq aria 1}.

\begin{table}[!htbp]
  \centering
  \caption{\textit{Quantinuum H1-1} calibration data (obtained April. 10, 2024).}\label{table:quantinuum H1-1}
  \begin{tabular}{c c}
    \hline\hline
      Relaxation time $T_1$ & $\gg$ 1 minute \\[0.2em]
      Decoherence time $T_2$ & $\approx$4 seconds \\[0.2em]
      SPAM error & 2.5$\times10^{-3}$ \\[0.2em]
      Single-qubit gate error & 2.1$\times10^{-5}$ \\[0.2em]
      Two-qubit gate error & 8.8$\times10^{-4}$ \\[0.2em]
    
    \hline\hline
  \end{tabular}
\end{table}

\subsection{Quantinuum}
For our demonstrations on Quantinuum, we used the backend $\textit{Quantinuum H1-1}$ which contains 20 trapped ion qubits with all-to-all connectivity. The calibration data of the device is shown in Table~\ref{table:quantinuum H1-1}.

% \bibliography{ref.bib}

\begin{thebibliography}{71}%
\makeatletter
\providecommand \@ifxundefined [1]{%
 \@ifx{#1\undefined}
}%
\providecommand \@ifnum [1]{%
 \ifnum #1\expandafter \@firstoftwo
 \else \expandafter \@secondoftwo
 \fi
}%
\providecommand \@ifx [1]{%
 \ifx #1\expandafter \@firstoftwo
 \else \expandafter \@secondoftwo
 \fi
}%
\providecommand \natexlab [1]{#1}%
\providecommand \enquote  [1]{``#1''}%
\providecommand \bibnamefont  [1]{#1}%
\providecommand \bibfnamefont [1]{#1}%
\providecommand \citenamefont [1]{#1}%
\providecommand \href@noop [0]{\@secondoftwo}%
\providecommand \href [0]{\begingroup \@sanitize@url \@href}%
\providecommand \@href[1]{\@@startlink{#1}\@@href}%
\providecommand \@@href[1]{\endgroup#1\@@endlink}%
\providecommand \@sanitize@url [0]{\catcode `\\12\catcode `\$12\catcode `\&12\catcode `\#12\catcode `\^12\catcode `\_12\catcode `\%12\relax}%
\providecommand \@@startlink[1]{}%
\providecommand \@@endlink[0]{}%
\providecommand \url  [0]{\begingroup\@sanitize@url \@url }%
\providecommand \@url [1]{\endgroup\@href {#1}{\urlprefix }}%
\providecommand \urlprefix  [0]{URL }%
\providecommand \Eprint [0]{\href }%
\providecommand \doibase [0]{https://doi.org/}%
\providecommand \selectlanguage [0]{\@gobble}%
\providecommand \bibinfo  [0]{\@secondoftwo}%
\providecommand \bibfield  [0]{\@secondoftwo}%
\providecommand \translation [1]{[#1]}%
\providecommand \BibitemOpen [0]{}%
\providecommand \bibitemStop [0]{}%
\providecommand \bibitemNoStop [0]{.\EOS\space}%
\providecommand \EOS [0]{\spacefactor3000\relax}%
\providecommand \BibitemShut  [1]{\csname bibitem#1\endcsname}%
\let\auto@bib@innerbib\@empty
%</preamble>
\bibitem [{\citenamefont {Campaioli}\ \emph {et~al.}(2024)\citenamefont {Campaioli}, \citenamefont {Gherardini}, \citenamefont {Quach}, \citenamefont {Polini},\ and\ \citenamefont {Andolina}}]{Campaioli2024}%
  \BibitemOpen
  \bibfield  {author} {\bibinfo {author} {\bibfnamefont {F.}~\bibnamefont {Campaioli}}, \bibinfo {author} {\bibfnamefont {S.}~\bibnamefont {Gherardini}}, \bibinfo {author} {\bibfnamefont {J.~Q.}\ \bibnamefont {Quach}}, \bibinfo {author} {\bibfnamefont {M.}~\bibnamefont {Polini}},\ and\ \bibinfo {author} {\bibfnamefont {G.~M.}\ \bibnamefont {Andolina}},\ }\bibfield  {title} {\bibinfo {title} {Colloquium: Quantum batteries},\ }\href {https://doi.org/10.1103/RevModPhys.96.031001} {\bibfield  {journal} {\bibinfo  {journal} {Rev. Mod. Phys.}\ }\textbf {\bibinfo {volume} {96}},\ \bibinfo {pages} {031001} (\bibinfo {year} {2024})}\BibitemShut {NoStop}%
\bibitem [{\citenamefont {Alicki}\ and\ \citenamefont {Fannes}(2013)}]{Alicki2013}%
  \BibitemOpen
  \bibfield  {author} {\bibinfo {author} {\bibfnamefont {R.}~\bibnamefont {Alicki}}\ and\ \bibinfo {author} {\bibfnamefont {M.}~\bibnamefont {Fannes}},\ }\bibfield  {title} {\bibinfo {title} {Entanglement boost for extractable work from ensembles of quantum batteries},\ }\href {https://doi.org/10.1103/PhysRevE.87.042123} {\bibfield  {journal} {\bibinfo  {journal} {Phys. Rev. E}\ }\textbf {\bibinfo {volume} {87}},\ \bibinfo {pages} {042123} (\bibinfo {year} {2013})}\BibitemShut {NoStop}%
\bibitem [{\citenamefont {Binder}\ \emph {et~al.}(2015)\citenamefont {Binder}, \citenamefont {Vinjanampathy}, \citenamefont {Modi},\ and\ \citenamefont {Goold}}]{Binder2015}%
  \BibitemOpen
  \bibfield  {author} {\bibinfo {author} {\bibfnamefont {F.~C.}\ \bibnamefont {Binder}}, \bibinfo {author} {\bibfnamefont {S.}~\bibnamefont {Vinjanampathy}}, \bibinfo {author} {\bibfnamefont {K.}~\bibnamefont {Modi}},\ and\ \bibinfo {author} {\bibfnamefont {J.}~\bibnamefont {Goold}},\ }\bibfield  {title} {\bibinfo {title} {Quantacell: powerful charging of quantum batteries},\ }\href {https://doi.org/10.1088/1367-2630/17/7/075015} {\bibfield  {journal} {\bibinfo  {journal} {New J. Phys.}\ }\textbf {\bibinfo {volume} {17}},\ \bibinfo {pages} {075015} (\bibinfo {year} {2015})}\BibitemShut {NoStop}%
\bibitem [{\citenamefont {Gemme}\ \emph {et~al.}(2022)\citenamefont {Gemme}, \citenamefont {Grossi}, \citenamefont {Ferraro}, \citenamefont {Vallecorsa},\ and\ \citenamefont {Sassetti}}]{Gemme2022}%
  \BibitemOpen
  \bibfield  {author} {\bibinfo {author} {\bibfnamefont {G.}~\bibnamefont {Gemme}}, \bibinfo {author} {\bibfnamefont {M.}~\bibnamefont {Grossi}}, \bibinfo {author} {\bibfnamefont {D.}~\bibnamefont {Ferraro}}, \bibinfo {author} {\bibfnamefont {S.}~\bibnamefont {Vallecorsa}},\ and\ \bibinfo {author} {\bibfnamefont {M.}~\bibnamefont {Sassetti}},\ }\bibfield  {title} {\bibinfo {title} {{IBM} quantum platforms: A quantum battery perspective},\ }\href {https://doi.org/10.3390/batteries8050043} {\bibfield  {journal} {\bibinfo  {journal} {Batteries}\ }\textbf {\bibinfo {volume} {8}},\ \bibinfo {pages} {43} (\bibinfo {year} {2022})}\BibitemShut {NoStop}%
\bibitem [{\citenamefont {Shi}\ \emph {et~al.}(2022)\citenamefont {Shi}, \citenamefont {Ding}, \citenamefont {Wan}, \citenamefont {Wang},\ and\ \citenamefont {Yang}}]{Shi2022}%
  \BibitemOpen
  \bibfield  {author} {\bibinfo {author} {\bibfnamefont {H.-L.}\ \bibnamefont {Shi}}, \bibinfo {author} {\bibfnamefont {S.}~\bibnamefont {Ding}}, \bibinfo {author} {\bibfnamefont {Q.-K.}\ \bibnamefont {Wan}}, \bibinfo {author} {\bibfnamefont {X.-H.}\ \bibnamefont {Wang}},\ and\ \bibinfo {author} {\bibfnamefont {W.-L.}\ \bibnamefont {Yang}},\ }\bibfield  {title} {\bibinfo {title} {{Entanglement, Coherence, and Extractable Work in Quantum Batteries}},\ }\href {https://doi.org/10.1103/PhysRevLett.129.130602} {\bibfield  {journal} {\bibinfo  {journal} {Phys. Rev. Lett.}\ }\textbf {\bibinfo {volume} {129}},\ \bibinfo {pages} {130602} (\bibinfo {year} {2022})}\BibitemShut {NoStop}%
\bibitem [{\citenamefont {Campaioli}\ \emph {et~al.}(2017)\citenamefont {Campaioli}, \citenamefont {Pollock}, \citenamefont {Binder}, \citenamefont {C\'eleri}, \citenamefont {Goold}, \citenamefont {Vinjanampathy},\ and\ \citenamefont {Modi}}]{Campaioli2017}%
  \BibitemOpen
  \bibfield  {author} {\bibinfo {author} {\bibfnamefont {F.}~\bibnamefont {Campaioli}}, \bibinfo {author} {\bibfnamefont {F.~A.}\ \bibnamefont {Pollock}}, \bibinfo {author} {\bibfnamefont {F.~C.}\ \bibnamefont {Binder}}, \bibinfo {author} {\bibfnamefont {L.}~\bibnamefont {C\'eleri}}, \bibinfo {author} {\bibfnamefont {J.}~\bibnamefont {Goold}}, \bibinfo {author} {\bibfnamefont {S.}~\bibnamefont {Vinjanampathy}},\ and\ \bibinfo {author} {\bibfnamefont {K.}~\bibnamefont {Modi}},\ }\bibfield  {title} {\bibinfo {title} {{Enhancing the Charging Power of Quantum Batteries}},\ }\href {https://doi.org/10.1103/PhysRevLett.118.150601} {\bibfield  {journal} {\bibinfo  {journal} {Phys. Rev. Lett.}\ }\textbf {\bibinfo {volume} {118}},\ \bibinfo {pages} {150601} (\bibinfo {year} {2017})}\BibitemShut {NoStop}%
\bibitem [{\citenamefont {Ferraro}\ \emph {et~al.}(2018)\citenamefont {Ferraro}, \citenamefont {Campisi}, \citenamefont {Andolina}, \citenamefont {Pellegrini},\ and\ \citenamefont {Polini}}]{Ferraro2018}%
  \BibitemOpen
  \bibfield  {author} {\bibinfo {author} {\bibfnamefont {D.}~\bibnamefont {Ferraro}}, \bibinfo {author} {\bibfnamefont {M.}~\bibnamefont {Campisi}}, \bibinfo {author} {\bibfnamefont {G.~M.}\ \bibnamefont {Andolina}}, \bibinfo {author} {\bibfnamefont {V.}~\bibnamefont {Pellegrini}},\ and\ \bibinfo {author} {\bibfnamefont {M.}~\bibnamefont {Polini}},\ }\bibfield  {title} {\bibinfo {title} {{High-Power Collective Charging of a Solid-State Quantum Battery}},\ }\href {https://doi.org/10.1103/PhysRevLett.120.117702} {\bibfield  {journal} {\bibinfo  {journal} {Phys. Rev. Lett.}\ }\textbf {\bibinfo {volume} {120}},\ \bibinfo {pages} {117702} (\bibinfo {year} {2018})}\BibitemShut {NoStop}%
\bibitem [{\citenamefont {Crescente}\ \emph {et~al.}(2020)\citenamefont {Crescente}, \citenamefont {Carrega}, \citenamefont {Sasetti},\ and\ \citenamefont {Ferraro}}]{Crescente2020}%
  \BibitemOpen
  \bibfield  {author} {\bibinfo {author} {\bibfnamefont {A.}~\bibnamefont {Crescente}}, \bibinfo {author} {\bibfnamefont {M.}~\bibnamefont {Carrega}}, \bibinfo {author} {\bibfnamefont {M.}~\bibnamefont {Sasetti}},\ and\ \bibinfo {author} {\bibfnamefont {D.}~\bibnamefont {Ferraro}},\ }\bibfield  {title} {\bibinfo {title} {Charging and energy fluctuations of a driven quantum battery},\ }\href {https://doi.org/10.1088/1367-2630/ab91fc} {\bibfield  {journal} {\bibinfo  {journal} {New J. Phys.}\ }\textbf {\bibinfo {volume} {22}},\ \bibinfo {pages} {063057} (\bibinfo {year} {2020})}\BibitemShut {NoStop}%
\bibitem [{\citenamefont {Ghosh}\ \emph {et~al.}(2020)\citenamefont {Ghosh}, \citenamefont {Chanda},\ and\ \citenamefont {Sen(De)}}]{Ghosh2020}%
  \BibitemOpen
  \bibfield  {author} {\bibinfo {author} {\bibfnamefont {S.}~\bibnamefont {Ghosh}}, \bibinfo {author} {\bibfnamefont {T.}~\bibnamefont {Chanda}},\ and\ \bibinfo {author} {\bibfnamefont {A.}~\bibnamefont {Sen(De)}},\ }\bibfield  {title} {\bibinfo {title} {Enhancement in the performance of a quantum battery by ordered and disordered interactions},\ }\href {https://doi.org/10.1103/PhysRevA.101.032115} {\bibfield  {journal} {\bibinfo  {journal} {Phys. Rev. A}\ }\textbf {\bibinfo {volume} {101}},\ \bibinfo {pages} {032115} (\bibinfo {year} {2020})}\BibitemShut {NoStop}%
\bibitem [{\citenamefont {Ghosh}\ \emph {et~al.}(2021)\citenamefont {Ghosh}, \citenamefont {Chanda}, \citenamefont {Mal},\ and\ \citenamefont {Sen(De)}}]{Ghosh2021}%
  \BibitemOpen
  \bibfield  {author} {\bibinfo {author} {\bibfnamefont {S.}~\bibnamefont {Ghosh}}, \bibinfo {author} {\bibfnamefont {T.}~\bibnamefont {Chanda}}, \bibinfo {author} {\bibfnamefont {S.}~\bibnamefont {Mal}},\ and\ \bibinfo {author} {\bibfnamefont {A.}~\bibnamefont {Sen(De)}},\ }\bibfield  {title} {\bibinfo {title} {Fast charging of a quantum battery assisted by noise},\ }\href {https://doi.org/10.1103/PhysRevA.104.032207} {\bibfield  {journal} {\bibinfo  {journal} {Phys. Rev. A}\ }\textbf {\bibinfo {volume} {104}},\ \bibinfo {pages} {032207} (\bibinfo {year} {2021})}\BibitemShut {NoStop}%
\bibitem [{\citenamefont {Quach}\ \emph {et~al.}(2022)\citenamefont {Quach}, \citenamefont {McGhee}, \citenamefont {Ganzer}, \citenamefont {Rouse}, \citenamefont {Lovett}, \citenamefont {Gauger}, \citenamefont {Keeling}, \citenamefont {Cerullo}, \citenamefont {Lidzey},\ and\ \citenamefont {Virgili}}]{Quach2022}%
  \BibitemOpen
  \bibfield  {author} {\bibinfo {author} {\bibfnamefont {J.~Q.}\ \bibnamefont {Quach}}, \bibinfo {author} {\bibfnamefont {K.~E.}\ \bibnamefont {McGhee}}, \bibinfo {author} {\bibfnamefont {L.}~\bibnamefont {Ganzer}}, \bibinfo {author} {\bibfnamefont {D.~M.}\ \bibnamefont {Rouse}}, \bibinfo {author} {\bibfnamefont {B.~W.}\ \bibnamefont {Lovett}}, \bibinfo {author} {\bibfnamefont {E.~M.}\ \bibnamefont {Gauger}}, \bibinfo {author} {\bibfnamefont {J.}~\bibnamefont {Keeling}}, \bibinfo {author} {\bibfnamefont {G.}~\bibnamefont {Cerullo}}, \bibinfo {author} {\bibfnamefont {D.~G.}\ \bibnamefont {Lidzey}},\ and\ \bibinfo {author} {\bibfnamefont {T.}~\bibnamefont {Virgili}},\ }\bibfield  {title} {\bibinfo {title} {Superabsorption in an organic microcavity: Toward a quantum battery},\ }\href {https://doi.org/10.1126/sciadv.abk3160} {\bibfield  {journal} {\bibinfo  {journal} {Sci. Adv.}\ }\textbf {\bibinfo {volume} {8}},\ \bibinfo {pages} {eabk3160} (\bibinfo {year} {2022})}\BibitemShut {NoStop}%
\bibitem [{\citenamefont {Ueki}\ \emph {et~al.}(2022)\citenamefont {Ueki}, \citenamefont {Kamimura}, \citenamefont {Matsuzaki}, \citenamefont {Yoshida},\ and\ \citenamefont {Tokura}}]{Ueki2022}%
  \BibitemOpen
  \bibfield  {author} {\bibinfo {author} {\bibfnamefont {Y.}~\bibnamefont {Ueki}}, \bibinfo {author} {\bibfnamefont {S.}~\bibnamefont {Kamimura}}, \bibinfo {author} {\bibfnamefont {Y.}~\bibnamefont {Matsuzaki}}, \bibinfo {author} {\bibfnamefont {K.}~\bibnamefont {Yoshida}},\ and\ \bibinfo {author} {\bibfnamefont {Y.}~\bibnamefont {Tokura}},\ }\bibfield  {title} {\bibinfo {title} {{Quantum Battery Based on Superabsorption}},\ }\href {https://doi.org/10.7566/jpsj.91.124002} {\bibfield  {journal} {\bibinfo  {journal} {J. Phys. Soc. Jpn.}\ }\textbf {\bibinfo {volume} {91}},\ \bibinfo {pages} {124002} (\bibinfo {year} {2022})}\BibitemShut {NoStop}%
\bibitem [{\citenamefont {Gemme}\ \emph {et~al.}(2023)\citenamefont {Gemme}, \citenamefont {Andolina}, \citenamefont {Pellegrino}, \citenamefont {Sassetti},\ and\ \citenamefont {Ferraro}}]{Gemme2023}%
  \BibitemOpen
  \bibfield  {author} {\bibinfo {author} {\bibfnamefont {G.}~\bibnamefont {Gemme}}, \bibinfo {author} {\bibfnamefont {G.~M.}\ \bibnamefont {Andolina}}, \bibinfo {author} {\bibfnamefont {F.~M.~D.}\ \bibnamefont {Pellegrino}}, \bibinfo {author} {\bibfnamefont {M.}~\bibnamefont {Sassetti}},\ and\ \bibinfo {author} {\bibfnamefont {D.}~\bibnamefont {Ferraro}},\ }\bibfield  {title} {\bibinfo {title} {{Off-resonant Dicke Quantum Battery: Charging by Virtual Photons}},\ }\href {https://doi.org/10.3390/batteries9040197} {\bibfield  {journal} {\bibinfo  {journal} {Batteries}\ }\textbf {\bibinfo {volume} {9}},\ \bibinfo {pages} {197} (\bibinfo {year} {2023})}\BibitemShut {NoStop}%
\bibitem [{\citenamefont {Andolina}\ \emph {et~al.}(2019)\citenamefont {Andolina}, \citenamefont {Keck}, \citenamefont {Mari}, \citenamefont {Campisi}, \citenamefont {Giovannetti},\ and\ \citenamefont {Polini}}]{Andolina2019}%
  \BibitemOpen
  \bibfield  {author} {\bibinfo {author} {\bibfnamefont {G.~M.}\ \bibnamefont {Andolina}}, \bibinfo {author} {\bibfnamefont {M.}~\bibnamefont {Keck}}, \bibinfo {author} {\bibfnamefont {A.}~\bibnamefont {Mari}}, \bibinfo {author} {\bibfnamefont {M.}~\bibnamefont {Campisi}}, \bibinfo {author} {\bibfnamefont {V.}~\bibnamefont {Giovannetti}},\ and\ \bibinfo {author} {\bibfnamefont {M.}~\bibnamefont {Polini}},\ }\bibfield  {title} {\bibinfo {title} {{Extractable Work, the Role of Correlations, and Asymptotic Freedom in Quantum Batteries}},\ }\href {https://doi.org/10.1103/PhysRevLett.122.047702} {\bibfield  {journal} {\bibinfo  {journal} {Phys. Rev. Lett.}\ }\textbf {\bibinfo {volume} {122}},\ \bibinfo {pages} {047702} (\bibinfo {year} {2019})}\BibitemShut {NoStop}%
\bibitem [{\citenamefont {Barra}(2019)}]{Barra2019}%
  \BibitemOpen
  \bibfield  {author} {\bibinfo {author} {\bibfnamefont {F.}~\bibnamefont {Barra}},\ }\bibfield  {title} {\bibinfo {title} {{Dissipative Charging of a Quantum Battery}},\ }\href {https://doi.org/10.1103/PhysRevLett.122.210601} {\bibfield  {journal} {\bibinfo  {journal} {Phys. Rev. Lett.}\ }\textbf {\bibinfo {volume} {122}},\ \bibinfo {pages} {210601} (\bibinfo {year} {2019})}\BibitemShut {NoStop}%
\bibitem [{\citenamefont {Kamin}\ \emph {et~al.}(2020)\citenamefont {Kamin}, \citenamefont {Tabesh}, \citenamefont {Salimi}, \citenamefont {Kheirandish},\ and\ \citenamefont {Santos}}]{Kamin2020-2}%
  \BibitemOpen
  \bibfield  {author} {\bibinfo {author} {\bibfnamefont {F.~H.}\ \bibnamefont {Kamin}}, \bibinfo {author} {\bibfnamefont {F.~T.}\ \bibnamefont {Tabesh}}, \bibinfo {author} {\bibfnamefont {S.}~\bibnamefont {Salimi}}, \bibinfo {author} {\bibfnamefont {F.}~\bibnamefont {Kheirandish}},\ and\ \bibinfo {author} {\bibfnamefont {A.~C.}\ \bibnamefont {Santos}},\ }\bibfield  {title} {\bibinfo {title} {Non-{M}arkovian effects on charging and self-discharging process of quantum batteries},\ }\href {https://doi.org/10.1088/1367-2630/ab9ee2} {\bibfield  {journal} {\bibinfo  {journal} {New J. Phys.}\ }\textbf {\bibinfo {volume} {22}},\ \bibinfo {pages} {083007} (\bibinfo {year} {2020})}\BibitemShut {NoStop}%
\bibitem [{\citenamefont {Francica}\ \emph {et~al.}(2020)\citenamefont {Francica}, \citenamefont {Binder}, \citenamefont {Guarnieri}, \citenamefont {Mitchison}, \citenamefont {Goold},\ and\ \citenamefont {Plastina}}]{Francica2020}%
  \BibitemOpen
  \bibfield  {author} {\bibinfo {author} {\bibfnamefont {G.}~\bibnamefont {Francica}}, \bibinfo {author} {\bibfnamefont {F.~C.}\ \bibnamefont {Binder}}, \bibinfo {author} {\bibfnamefont {G.}~\bibnamefont {Guarnieri}}, \bibinfo {author} {\bibfnamefont {M.~T.}\ \bibnamefont {Mitchison}}, \bibinfo {author} {\bibfnamefont {J.}~\bibnamefont {Goold}},\ and\ \bibinfo {author} {\bibfnamefont {F.}~\bibnamefont {Plastina}},\ }\bibfield  {title} {\bibinfo {title} {{Quantum Coherence and Ergotropy}},\ }\href {https://doi.org/10.1103/PhysRevLett.125.180603} {\bibfield  {journal} {\bibinfo  {journal} {Phys. Rev. Lett.}\ }\textbf {\bibinfo {volume} {125}},\ \bibinfo {pages} {180603} (\bibinfo {year} {2020})}\BibitemShut {NoStop}%
\bibitem [{\citenamefont {Lai}\ \emph {et~al.}(2024)\citenamefont {Lai}, \citenamefont {Lin}, \citenamefont {Huang}, \citenamefont {Jan},\ and\ \citenamefont {Chen}}]{Lai2024}%
  \BibitemOpen
  \bibfield  {author} {\bibinfo {author} {\bibfnamefont {P.-R.}\ \bibnamefont {Lai}}, \bibinfo {author} {\bibfnamefont {J.-D.}\ \bibnamefont {Lin}}, \bibinfo {author} {\bibfnamefont {Y.-T.}\ \bibnamefont {Huang}}, \bibinfo {author} {\bibfnamefont {H.-C.}\ \bibnamefont {Jan}},\ and\ \bibinfo {author} {\bibfnamefont {Y.-N.}\ \bibnamefont {Chen}},\ }\bibfield  {title} {\bibinfo {title} {Quick charging of a quantum battery with superposed trajectories},\ }\href {https://doi.org/10.1103/PhysRevResearch.6.023136} {\bibfield  {journal} {\bibinfo  {journal} {Phys. Rev. Res.}\ }\textbf {\bibinfo {volume} {6}},\ \bibinfo {pages} {023136} (\bibinfo {year} {2024})}\BibitemShut {NoStop}%
\bibitem [{\citenamefont {Le}\ \emph {et~al.}(2018)\citenamefont {Le}, \citenamefont {Levinsen}, \citenamefont {Modi}, \citenamefont {Parish},\ and\ \citenamefont {Pollock}}]{Le2018}%
  \BibitemOpen
  \bibfield  {author} {\bibinfo {author} {\bibfnamefont {T.~P.}\ \bibnamefont {Le}}, \bibinfo {author} {\bibfnamefont {J.}~\bibnamefont {Levinsen}}, \bibinfo {author} {\bibfnamefont {K.}~\bibnamefont {Modi}}, \bibinfo {author} {\bibfnamefont {M.~M.}\ \bibnamefont {Parish}},\ and\ \bibinfo {author} {\bibfnamefont {F.~A.}\ \bibnamefont {Pollock}},\ }\bibfield  {title} {\bibinfo {title} {Spin-chain model of a many-body quantum battery},\ }\href {https://doi.org/10.1103/PhysRevA.97.022106} {\bibfield  {journal} {\bibinfo  {journal} {Phys. Rev. A}\ }\textbf {\bibinfo {volume} {97}},\ \bibinfo {pages} {022106} (\bibinfo {year} {2018})}\BibitemShut {NoStop}%
\bibitem [{\citenamefont {Manzano}\ \emph {et~al.}(2018)\citenamefont {Manzano}, \citenamefont {Plastina},\ and\ \citenamefont {Zambrini}}]{Manzano2018}%
  \BibitemOpen
  \bibfield  {author} {\bibinfo {author} {\bibfnamefont {G.}~\bibnamefont {Manzano}}, \bibinfo {author} {\bibfnamefont {F.}~\bibnamefont {Plastina}},\ and\ \bibinfo {author} {\bibfnamefont {R.}~\bibnamefont {Zambrini}},\ }\bibfield  {title} {\bibinfo {title} {{Optimal Work Extraction and Thermodynamics of Quantum Measurements and Correlations}},\ }\href {https://doi.org/10.1103/PhysRevLett.121.120602} {\bibfield  {journal} {\bibinfo  {journal} {Phys. Rev. Lett.}\ }\textbf {\bibinfo {volume} {121}},\ \bibinfo {pages} {120602} (\bibinfo {year} {2018})}\BibitemShut {NoStop}%
\bibitem [{\citenamefont {Rossini}\ \emph {et~al.}(2019)\citenamefont {Rossini}, \citenamefont {Andolina},\ and\ \citenamefont {Polini}}]{Rossini2019}%
  \BibitemOpen
  \bibfield  {author} {\bibinfo {author} {\bibfnamefont {D.}~\bibnamefont {Rossini}}, \bibinfo {author} {\bibfnamefont {G.~M.}\ \bibnamefont {Andolina}},\ and\ \bibinfo {author} {\bibfnamefont {M.}~\bibnamefont {Polini}},\ }\bibfield  {title} {\bibinfo {title} {Many-body localized quantum batteries},\ }\href {https://doi.org/10.1103/PhysRevB.100.115142} {\bibfield  {journal} {\bibinfo  {journal} {Phys. Rev. B}\ }\textbf {\bibinfo {volume} {100}},\ \bibinfo {pages} {115142} (\bibinfo {year} {2019})}\BibitemShut {NoStop}%
\bibitem [{\citenamefont {Santos}\ \emph {et~al.}(2019)\citenamefont {Santos}, \citenamefont {\c{C}akmak}, \citenamefont {Campbell},\ and\ \citenamefont {Zinner}}]{Santos2019}%
  \BibitemOpen
  \bibfield  {author} {\bibinfo {author} {\bibfnamefont {A.~C.}\ \bibnamefont {Santos}}, \bibinfo {author} {\bibfnamefont {B.}~\bibnamefont {\c{C}akmak}}, \bibinfo {author} {\bibfnamefont {S.}~\bibnamefont {Campbell}},\ and\ \bibinfo {author} {\bibfnamefont {N.~T.}\ \bibnamefont {Zinner}},\ }\bibfield  {title} {\bibinfo {title} {Stable adiabatic quantum batteries},\ }\href {https://doi.org/10.1103/PhysRevE.100.032107} {\bibfield  {journal} {\bibinfo  {journal} {Phys. Rev. E}\ }\textbf {\bibinfo {volume} {100}},\ \bibinfo {pages} {032107} (\bibinfo {year} {2019})}\BibitemShut {NoStop}%
\bibitem [{\citenamefont {Pirmoradian}\ and\ \citenamefont {M\o{}lmer}(2019)}]{Primoradian2019}%
  \BibitemOpen
  \bibfield  {author} {\bibinfo {author} {\bibfnamefont {F.}~\bibnamefont {Pirmoradian}}\ and\ \bibinfo {author} {\bibfnamefont {K.}~\bibnamefont {M\o{}lmer}},\ }\bibfield  {title} {\bibinfo {title} {Aging of a quantum battery},\ }\href {https://doi.org/10.1103/PhysRevA.100.043833} {\bibfield  {journal} {\bibinfo  {journal} {Phys. Rev. A}\ }\textbf {\bibinfo {volume} {100}},\ \bibinfo {pages} {043833} (\bibinfo {year} {2019})}\BibitemShut {NoStop}%
\bibitem [{\citenamefont {Liu}\ \emph {et~al.}(2019)\citenamefont {Liu}, \citenamefont {Segal},\ and\ \citenamefont {Hanna}}]{Liu2019}%
  \BibitemOpen
  \bibfield  {author} {\bibinfo {author} {\bibfnamefont {J.}~\bibnamefont {Liu}}, \bibinfo {author} {\bibfnamefont {D.}~\bibnamefont {Segal}},\ and\ \bibinfo {author} {\bibfnamefont {G.}~\bibnamefont {Hanna}},\ }\bibfield  {title} {\bibinfo {title} {{Loss-Free Excitonic Quantum Battery}},\ }\href {https://doi.org/doi: 10.1021/acs.jpcc.9b06373} {\bibfield  {journal} {\bibinfo  {journal} {J. Phys. Chem. C}\ }\textbf {\bibinfo {volume} {123}},\ \bibinfo {pages} {18303} (\bibinfo {year} {2019})}\BibitemShut {NoStop}%
\bibitem [{\citenamefont {Rossini}\ \emph {et~al.}(2020)\citenamefont {Rossini}, \citenamefont {Andolina}, \citenamefont {Rosa}, \citenamefont {Carrega},\ and\ \citenamefont {Polini}}]{Rossini2020}%
  \BibitemOpen
  \bibfield  {author} {\bibinfo {author} {\bibfnamefont {D.}~\bibnamefont {Rossini}}, \bibinfo {author} {\bibfnamefont {G.~M.}\ \bibnamefont {Andolina}}, \bibinfo {author} {\bibfnamefont {D.}~\bibnamefont {Rosa}}, \bibinfo {author} {\bibfnamefont {M.}~\bibnamefont {Carrega}},\ and\ \bibinfo {author} {\bibfnamefont {M.}~\bibnamefont {Polini}},\ }\bibfield  {title} {\bibinfo {title} {{Quantum Advantage in the Charging Process of Sachdev-Ye-Kitaev Batteries}},\ }\href {https://doi.org/10.1103/PhysRevLett.125.236402} {\bibfield  {journal} {\bibinfo  {journal} {Phys. Rev. Lett.}\ }\textbf {\bibinfo {volume} {125}},\ \bibinfo {pages} {236402} (\bibinfo {year} {2020})}\BibitemShut {NoStop}%
\bibitem [{\citenamefont {Quach}\ and\ \citenamefont {Munro}(2020)}]{Quach2020}%
  \BibitemOpen
  \bibfield  {author} {\bibinfo {author} {\bibfnamefont {J.~Q.}\ \bibnamefont {Quach}}\ and\ \bibinfo {author} {\bibfnamefont {W.~J.}\ \bibnamefont {Munro}},\ }\bibfield  {title} {\bibinfo {title} {{Using Dark States to Charge and Stabilize Open Quantum Batteries}},\ }\href {https://doi.org/10.1103/PhysRevApplied.14.024092} {\bibfield  {journal} {\bibinfo  {journal} {Phys. Rev. Applied}\ }\textbf {\bibinfo {volume} {14}},\ \bibinfo {pages} {024092} (\bibinfo {year} {2020})}\BibitemShut {NoStop}%
\bibitem [{\citenamefont {Monsel}\ \emph {et~al.}(2020)\citenamefont {Monsel}, \citenamefont {Fellous-Asiani}, \citenamefont {Huard},\ and\ \citenamefont {Auff\`eves}}]{Monsel2020}%
  \BibitemOpen
  \bibfield  {author} {\bibinfo {author} {\bibfnamefont {J.}~\bibnamefont {Monsel}}, \bibinfo {author} {\bibfnamefont {M.}~\bibnamefont {Fellous-Asiani}}, \bibinfo {author} {\bibfnamefont {B.}~\bibnamefont {Huard}},\ and\ \bibinfo {author} {\bibfnamefont {A.}~\bibnamefont {Auff\`eves}},\ }\bibfield  {title} {\bibinfo {title} {{The Energetic Cost of Work Extraction}},\ }\href {https://doi.org/10.1103/PhysRevLett.124.130601} {\bibfield  {journal} {\bibinfo  {journal} {Phys. Rev. Lett.}\ }\textbf {\bibinfo {volume} {124}},\ \bibinfo {pages} {130601} (\bibinfo {year} {2020})}\BibitemShut {NoStop}%
\bibitem [{\citenamefont {Chiribella}\ and\ \citenamefont {Kristj\'{a}nsson}(2019)}]{Chiribella2019}%
  \BibitemOpen
  \bibfield  {author} {\bibinfo {author} {\bibfnamefont {G.}~\bibnamefont {Chiribella}}\ and\ \bibinfo {author} {\bibfnamefont {H.}~\bibnamefont {Kristj\'{a}nsson}},\ }\bibfield  {title} {\bibinfo {title} {Quantum {S}hannon theory with superposition of trajectories},\ }\href {http://doi.org/10.1098/rspa.2018.0903} {\bibfield  {journal} {\bibinfo  {journal} {Proc. R. Soc. A}\ }\textbf {\bibinfo {volume} {475}},\ \bibinfo {pages} {20180903} (\bibinfo {year} {2019})}\BibitemShut {NoStop}%
\bibitem [{\citenamefont {Foo}\ \emph {et~al.}(2020)\citenamefont {Foo}, \citenamefont {Onoe},\ and\ \citenamefont {Zych}}]{Foo2020}%
  \BibitemOpen
  \bibfield  {author} {\bibinfo {author} {\bibfnamefont {J.}~\bibnamefont {Foo}}, \bibinfo {author} {\bibfnamefont {S.}~\bibnamefont {Onoe}},\ and\ \bibinfo {author} {\bibfnamefont {M.}~\bibnamefont {Zych}},\ }\bibfield  {title} {\bibinfo {title} {{U}nruh-{D}e{W}itt detectors in quantum superpositions of trajectories},\ }\href {https://doi.org/10.1103/PhysRevD.102.085013} {\bibfield  {journal} {\bibinfo  {journal} {Phys. Rev. D}\ }\textbf {\bibinfo {volume} {102}},\ \bibinfo {pages} {085013} (\bibinfo {year} {2020})}\BibitemShut {NoStop}%
\bibitem [{\citenamefont {Gisin}\ \emph {et~al.}(2005)\citenamefont {Gisin}, \citenamefont {Linden}, \citenamefont {Massar},\ and\ \citenamefont {Popescu}}]{Gisin2005}%
  \BibitemOpen
  \bibfield  {author} {\bibinfo {author} {\bibfnamefont {N.}~\bibnamefont {Gisin}}, \bibinfo {author} {\bibfnamefont {N.}~\bibnamefont {Linden}}, \bibinfo {author} {\bibfnamefont {S.}~\bibnamefont {Massar}},\ and\ \bibinfo {author} {\bibfnamefont {S.}~\bibnamefont {Popescu}},\ }\bibfield  {title} {\bibinfo {title} {Error filtration and entanglement purification for quantum communication},\ }\href {https://doi.org/10.1103/PhysRevA.72.012338} {\bibfield  {journal} {\bibinfo  {journal} {Phys. Rev. A}\ }\textbf {\bibinfo {volume} {72}},\ \bibinfo {pages} {012338} (\bibinfo {year} {2005})}\BibitemShut {NoStop}%
\bibitem [{\citenamefont {Kristj\'ansson}\ \emph {et~al.}(2020)\citenamefont {Kristj\'ansson}, \citenamefont {Chiribella}, \citenamefont {Salek}, \citenamefont {Ebler},\ and\ \citenamefont {Wilson}}]{Kristjansson2020}%
  \BibitemOpen
  \bibfield  {author} {\bibinfo {author} {\bibfnamefont {H.}~\bibnamefont {Kristj\'ansson}}, \bibinfo {author} {\bibfnamefont {G.}~\bibnamefont {Chiribella}}, \bibinfo {author} {\bibfnamefont {S.}~\bibnamefont {Salek}}, \bibinfo {author} {\bibfnamefont {D.}~\bibnamefont {Ebler}},\ and\ \bibinfo {author} {\bibfnamefont {M.}~\bibnamefont {Wilson}},\ }\bibfield  {title} {\bibinfo {title} {Resource theories of communication},\ }\href {https://doi.org/10.1088/1367-2630/ab8ef7} {\bibfield  {journal} {\bibinfo  {journal} {New J. Phys.}\ }\textbf {\bibinfo {volume} {22}} (\bibinfo {year} {2020})}\BibitemShut {NoStop}%
\bibitem [{\citenamefont {Rubino}\ \emph {et~al.}(2021)\citenamefont {Rubino}, \citenamefont {Rozema}, \citenamefont {Ebler}, \citenamefont {Kristj\'ansson}, \citenamefont {Salek}, \citenamefont {Allard~Gu\'erin}, \citenamefont {Abbott}, \citenamefont {Branciard}, \citenamefont {Brukner}, \citenamefont {Chiribella},\ and\ \citenamefont {Walther}}]{Rubino2021}%
  \BibitemOpen
  \bibfield  {author} {\bibinfo {author} {\bibfnamefont {G.}~\bibnamefont {Rubino}}, \bibinfo {author} {\bibfnamefont {L.~A.}\ \bibnamefont {Rozema}}, \bibinfo {author} {\bibfnamefont {D.}~\bibnamefont {Ebler}}, \bibinfo {author} {\bibfnamefont {H.}~\bibnamefont {Kristj\'ansson}}, \bibinfo {author} {\bibfnamefont {S.}~\bibnamefont {Salek}}, \bibinfo {author} {\bibfnamefont {P.}~\bibnamefont {Allard~Gu\'erin}}, \bibinfo {author} {\bibfnamefont {A.~A.}\ \bibnamefont {Abbott}}, \bibinfo {author} {\bibfnamefont {C.}~\bibnamefont {Branciard}}, \bibinfo {author} {\bibfnamefont {{\v{C}}.}~\bibnamefont {Brukner}}, \bibinfo {author} {\bibfnamefont {G.}~\bibnamefont {Chiribella}},\ and\ \bibinfo {author} {\bibfnamefont {P.}~\bibnamefont {Walther}},\ }\bibfield  {title} {\bibinfo {title} {Experimental quantum communication enhancement by superposing trajectories},\ }\href {https://doi.org/10.1103/PhysRevResearch.3.013093} {\bibfield  {journal} {\bibinfo  {journal} {Phys. Rev. Res.}\ }\textbf {\bibinfo {volume} {3}},\
  \bibinfo {pages} {013093} (\bibinfo {year} {2021})}\BibitemShut {NoStop}%
\bibitem [{\citenamefont {Ku}\ \emph {et~al.}(2023)\citenamefont {Ku}, \citenamefont {Lee}, \citenamefont {Lai}, \citenamefont {Lin},\ and\ \citenamefont {Chen}}]{Ku2023}%
  \BibitemOpen
  \bibfield  {author} {\bibinfo {author} {\bibfnamefont {H.-Y.}\ \bibnamefont {Ku}}, \bibinfo {author} {\bibfnamefont {K.-Y.}\ \bibnamefont {Lee}}, \bibinfo {author} {\bibfnamefont {P.-R.}\ \bibnamefont {Lai}}, \bibinfo {author} {\bibfnamefont {J.-D.}\ \bibnamefont {Lin}},\ and\ \bibinfo {author} {\bibfnamefont {Y.-N.}\ \bibnamefont {Chen}},\ }\bibfield  {title} {\bibinfo {title} {Coherent activation of a steerability-breaking channel},\ }\href {https://doi.org/10.1103/PhysRevA.107.042415} {\bibfield  {journal} {\bibinfo  {journal} {Phys. Rev. A}\ }\textbf {\bibinfo {volume} {107}},\ \bibinfo {pages} {042415} (\bibinfo {year} {2023})}\BibitemShut {NoStop}%
\bibitem [{\citenamefont {Ban}(2020)}]{Ban2020}%
  \BibitemOpen
  \bibfield  {author} {\bibinfo {author} {\bibfnamefont {M.}~\bibnamefont {Ban}},\ }\bibfield  {title} {\bibinfo {title} {Relaxation process of a two-level system in a coherent superposition of two environments},\ }\href {https://doi.org/10.1007/s11128-020-02856-6} {\bibfield  {journal} {\bibinfo  {journal} {Quantum Inf. Process.}\ }\textbf {\bibinfo {volume} {19}},\ \bibinfo {pages} {351} (\bibinfo {year} {2020})}\BibitemShut {NoStop}%
\bibitem [{\citenamefont {Chan}\ \emph {et~al.}(2022)\citenamefont {Chan}, \citenamefont {Huang}, \citenamefont {Lin}, \citenamefont {Ku}, \citenamefont {Chen}, \citenamefont {Chen},\ and\ \citenamefont {Chen}}]{Chan2022}%
  \BibitemOpen
  \bibfield  {author} {\bibinfo {author} {\bibfnamefont {F.-J.}\ \bibnamefont {Chan}}, \bibinfo {author} {\bibfnamefont {Y.-T.}\ \bibnamefont {Huang}}, \bibinfo {author} {\bibfnamefont {J.-D.}\ \bibnamefont {Lin}}, \bibinfo {author} {\bibfnamefont {H.-Y.}\ \bibnamefont {Ku}}, \bibinfo {author} {\bibfnamefont {J.-S.}\ \bibnamefont {Chen}}, \bibinfo {author} {\bibfnamefont {H.-B.}\ \bibnamefont {Chen}},\ and\ \bibinfo {author} {\bibfnamefont {Y.-N.}\ \bibnamefont {Chen}},\ }\bibfield  {title} {\bibinfo {title} {Maxwell's two-demon engine under pure dephasing noise},\ }\href {https://doi.org/10.1103/PhysRevA.106.052201} {\bibfield  {journal} {\bibinfo  {journal} {Phys. Rev. A}\ }\textbf {\bibinfo {volume} {106}},\ \bibinfo {pages} {052201} (\bibinfo {year} {2022})}\BibitemShut {NoStop}%
\bibitem [{\citenamefont {Lee}\ \emph {et~al.}(2023)\citenamefont {Lee}, \citenamefont {Lin}, \citenamefont {Miranowicz}, \citenamefont {Nori}, \citenamefont {Ku},\ and\ \citenamefont {Chen}}]{Lee2023}%
  \BibitemOpen
  \bibfield  {author} {\bibinfo {author} {\bibfnamefont {K.-Y.}\ \bibnamefont {Lee}}, \bibinfo {author} {\bibfnamefont {J.-D.}\ \bibnamefont {Lin}}, \bibinfo {author} {\bibfnamefont {A.}~\bibnamefont {Miranowicz}}, \bibinfo {author} {\bibfnamefont {F.}~\bibnamefont {Nori}}, \bibinfo {author} {\bibfnamefont {H.-Y.}\ \bibnamefont {Ku}},\ and\ \bibinfo {author} {\bibfnamefont {Y.-N.}\ \bibnamefont {Chen}},\ }\bibfield  {title} {\bibinfo {title} {Steering-enhanced quantum metrology using superpositions of noisy phase shifts},\ }\href {https://doi.org/10.1103/PhysRevResearch.5.013103} {\bibfield  {journal} {\bibinfo  {journal} {Phys. Rev. Res.}\ }\textbf {\bibinfo {volume} {5}},\ \bibinfo {pages} {013103} (\bibinfo {year} {2023})}\BibitemShut {NoStop}%
\bibitem [{\citenamefont {Foo}\ \emph {et~al.}(2021)\citenamefont {Foo}, \citenamefont {Mann},\ and\ \citenamefont {Zych}}]{Foo2021-2}%
  \BibitemOpen
  \bibfield  {author} {\bibinfo {author} {\bibfnamefont {J.}~\bibnamefont {Foo}}, \bibinfo {author} {\bibfnamefont {R.~B.}\ \bibnamefont {Mann}},\ and\ \bibinfo {author} {\bibfnamefont {M.}~\bibnamefont {Zych}},\ }\bibfield  {title} {\bibinfo {title} {Entanglement amplification between superposed detectors in flat and curved spacetimes},\ }\href {https://doi.org/10.1103/PhysRevD.103.065013} {\bibfield  {journal} {\bibinfo  {journal} {Phys. Rev. D}\ }\textbf {\bibinfo {volume} {103}},\ \bibinfo {pages} {065013} (\bibinfo {year} {2021})}\BibitemShut {NoStop}%
\bibitem [{\citenamefont {Siltanen}\ \emph {et~al.}(2021)\citenamefont {Siltanen}, \citenamefont {Kuusela},\ and\ \citenamefont {Piilo}}]{Siltanen2021}%
  \BibitemOpen
  \bibfield  {author} {\bibinfo {author} {\bibfnamefont {O.}~\bibnamefont {Siltanen}}, \bibinfo {author} {\bibfnamefont {T.}~\bibnamefont {Kuusela}},\ and\ \bibinfo {author} {\bibfnamefont {J.}~\bibnamefont {Piilo}},\ }\bibfield  {title} {\bibinfo {title} {Interferometric approach to open quantum systems and non-markovian dynamics},\ }\href {https://doi.org/10.1103/PhysRevA.103.032223} {\bibfield  {journal} {\bibinfo  {journal} {Phys. Rev. A}\ }\textbf {\bibinfo {volume} {103}},\ \bibinfo {pages} {032223} (\bibinfo {year} {2021})}\BibitemShut {NoStop}%
\bibitem [{\citenamefont {Lin}\ \emph {et~al.}(2022)\citenamefont {Lin}, \citenamefont {Huang}, \citenamefont {Lambert}, \citenamefont {Chen}, \citenamefont {Nori},\ and\ \citenamefont {Chen}}]{Lin2022}%
  \BibitemOpen
  \bibfield  {author} {\bibinfo {author} {\bibfnamefont {J.-D.}\ \bibnamefont {Lin}}, \bibinfo {author} {\bibfnamefont {C.-Y.}\ \bibnamefont {Huang}}, \bibinfo {author} {\bibfnamefont {N.}~\bibnamefont {Lambert}}, \bibinfo {author} {\bibfnamefont {G.-Y.}\ \bibnamefont {Chen}}, \bibinfo {author} {\bibfnamefont {F.}~\bibnamefont {Nori}},\ and\ \bibinfo {author} {\bibfnamefont {Y.-N.}\ \bibnamefont {Chen}},\ }\bibfield  {title} {\bibinfo {title} {Space-time dual quantum {Z}eno effect: {I}nterferometric engineering of open quantum system dynamics},\ }\href {https://doi.org/10.1103/PhysRevResearch.4.033143} {\bibfield  {journal} {\bibinfo  {journal} {Phys. Rev. Research}\ }\textbf {\bibinfo {volume} {4}},\ \bibinfo {pages} {033143} (\bibinfo {year} {2022})}\BibitemShut {NoStop}%
\bibitem [{\citenamefont {Lin}\ and\ \citenamefont {Chen}(2023)}]{Lin2023}%
  \BibitemOpen
  \bibfield  {author} {\bibinfo {author} {\bibfnamefont {J.-D.}\ \bibnamefont {Lin}}\ and\ \bibinfo {author} {\bibfnamefont {Y.-N.}\ \bibnamefont {Chen}},\ }\bibfield  {title} {\bibinfo {title} {Boosting entanglement growth of many-body localization by superpositions of disorder},\ }\href {https://doi.org/10.1103/PhysRevA.108.022203} {\bibfield  {journal} {\bibinfo  {journal} {Phys. Rev. A}\ }\textbf {\bibinfo {volume} {108}},\ \bibinfo {pages} {022203} (\bibinfo {year} {2023})}\BibitemShut {NoStop}%
\bibitem [{\citenamefont {\AA{}berg}(2014)}]{Aberg2014}%
  \BibitemOpen
  \bibfield  {author} {\bibinfo {author} {\bibfnamefont {J.}~\bibnamefont {\AA{}berg}},\ }\bibfield  {title} {\bibinfo {title} {{Catalytic Coherence}},\ }\href {https://doi.org/10.1103/PhysRevLett.113.150402} {\bibfield  {journal} {\bibinfo  {journal} {Phys. Rev. Lett.}\ }\textbf {\bibinfo {volume} {113}},\ \bibinfo {pages} {150402} (\bibinfo {year} {2014})}\BibitemShut {NoStop}%
\bibitem [{\citenamefont {Lostaglio}\ \emph {et~al.}(2015{\natexlab{a}})\citenamefont {Lostaglio}, \citenamefont {Korzekwa}, \citenamefont {Jennings},\ and\ \citenamefont {Rudolph}}]{Lostaglio2015}%
  \BibitemOpen
  \bibfield  {author} {\bibinfo {author} {\bibfnamefont {M.}~\bibnamefont {Lostaglio}}, \bibinfo {author} {\bibfnamefont {K.}~\bibnamefont {Korzekwa}}, \bibinfo {author} {\bibfnamefont {D.}~\bibnamefont {Jennings}},\ and\ \bibinfo {author} {\bibfnamefont {T.}~\bibnamefont {Rudolph}},\ }\bibfield  {title} {\bibinfo {title} {{Quantum Coherence, Time-Translation Symmetry, and Thermodynamics}},\ }\href {https://doi.org/10.1103/PhysRevX.5.021001} {\bibfield  {journal} {\bibinfo  {journal} {Phys. Rev. X}\ }\textbf {\bibinfo {volume} {5}},\ \bibinfo {pages} {021001} (\bibinfo {year} {2015}{\natexlab{a}})}\BibitemShut {NoStop}%
\bibitem [{\citenamefont {Lostaglio}\ \emph {et~al.}(2015{\natexlab{b}})\citenamefont {Lostaglio}, \citenamefont {Jennings},\ and\ \citenamefont {Rudolph}}]{Lostaglio2015_2}%
  \BibitemOpen
  \bibfield  {author} {\bibinfo {author} {\bibfnamefont {M.}~\bibnamefont {Lostaglio}}, \bibinfo {author} {\bibfnamefont {D.}~\bibnamefont {Jennings}},\ and\ \bibinfo {author} {\bibfnamefont {T.}~\bibnamefont {Rudolph}},\ }\bibfield  {title} {\bibinfo {title} {Description of quantum coherence in thermodynamic processes requires constraints beyond free energy},\ }\href {https://doi.org/10.1038/ncomms7383} {\bibfield  {journal} {\bibinfo  {journal} {Nat. Commun.}\ }\textbf {\bibinfo {volume} {6}},\ \bibinfo {pages} {6383} (\bibinfo {year} {2015}{\natexlab{b}})}\BibitemShut {NoStop}%
\bibitem [{\citenamefont {Rubino}\ \emph {et~al.}(2017)\citenamefont {Rubino}, \citenamefont {Rozema}, \citenamefont {Feix}, \citenamefont {Ara\'ujo}, \citenamefont {Zeuner}, \citenamefont {Procopio}, \citenamefont {Brukner},\ and\ \citenamefont {Walther}}]{Rubino2017}%
  \BibitemOpen
  \bibfield  {author} {\bibinfo {author} {\bibfnamefont {G.}~\bibnamefont {Rubino}}, \bibinfo {author} {\bibfnamefont {L.~A.}\ \bibnamefont {Rozema}}, \bibinfo {author} {\bibfnamefont {A.}~\bibnamefont {Feix}}, \bibinfo {author} {\bibfnamefont {M.}~\bibnamefont {Ara\'ujo}}, \bibinfo {author} {\bibfnamefont {J.~M.}\ \bibnamefont {Zeuner}}, \bibinfo {author} {\bibfnamefont {L.~M.}\ \bibnamefont {Procopio}}, \bibinfo {author} {\bibfnamefont {{\v{C}}.}~\bibnamefont {Brukner}},\ and\ \bibinfo {author} {\bibfnamefont {P.}~\bibnamefont {Walther}},\ }\bibfield  {title} {\bibinfo {title} {Experimental verification of an indefinite causal order},\ }\href {https://doi.org/10.1126/sciadv.1602589} {\bibfield  {journal} {\bibinfo  {journal} {Sci. Adv.}\ }\textbf {\bibinfo {volume} {3}},\ \bibinfo {pages} {e1602589} (\bibinfo {year} {2017})}\BibitemShut {NoStop}%
\bibitem [{\citenamefont {Goswami}\ \emph {et~al.}(2018)\citenamefont {Goswami}, \citenamefont {Giarmatzi}, \citenamefont {Kewming}, \citenamefont {Costa}, \citenamefont {Branciard}, \citenamefont {Romero},\ and\ \citenamefont {White}}]{Goswami2018}%
  \BibitemOpen
  \bibfield  {author} {\bibinfo {author} {\bibfnamefont {K.}~\bibnamefont {Goswami}}, \bibinfo {author} {\bibfnamefont {C.}~\bibnamefont {Giarmatzi}}, \bibinfo {author} {\bibfnamefont {M.}~\bibnamefont {Kewming}}, \bibinfo {author} {\bibfnamefont {F.}~\bibnamefont {Costa}}, \bibinfo {author} {\bibfnamefont {C.}~\bibnamefont {Branciard}}, \bibinfo {author} {\bibfnamefont {J.}~\bibnamefont {Romero}},\ and\ \bibinfo {author} {\bibfnamefont {A.~G.}\ \bibnamefont {White}},\ }\bibfield  {title} {\bibinfo {title} {{Indefinite Causal Order in a Quantum Switch}},\ }\href {https://doi.org/10.1103/PhysRevLett.121.090503} {\bibfield  {journal} {\bibinfo  {journal} {Phys. Rev. Lett.}\ }\textbf {\bibinfo {volume} {121}},\ \bibinfo {pages} {090503} (\bibinfo {year} {2018})}\BibitemShut {NoStop}%
\bibitem [{\citenamefont {Ebler}\ \emph {et~al.}(2018)\citenamefont {Ebler}, \citenamefont {Salek},\ and\ \citenamefont {Chiribella}}]{Ebler2018}%
  \BibitemOpen
  \bibfield  {author} {\bibinfo {author} {\bibfnamefont {D.}~\bibnamefont {Ebler}}, \bibinfo {author} {\bibfnamefont {S.}~\bibnamefont {Salek}},\ and\ \bibinfo {author} {\bibfnamefont {G.}~\bibnamefont {Chiribella}},\ }\bibfield  {title} {\bibinfo {title} {{Enhanced Communication with the Assistance of Indefinite Causal Order}},\ }\href {https://doi.org/10.1103/PhysRevLett.120.120502} {\bibfield  {journal} {\bibinfo  {journal} {Phys. Rev. Lett.}\ }\textbf {\bibinfo {volume} {120}},\ \bibinfo {pages} {120502} (\bibinfo {year} {2018})}\BibitemShut {NoStop}%
\bibitem [{\citenamefont {Zhao}\ \emph {et~al.}(2020)\citenamefont {Zhao}, \citenamefont {Yang},\ and\ \citenamefont {Chiribella}}]{Zhao2020}%
  \BibitemOpen
  \bibfield  {author} {\bibinfo {author} {\bibfnamefont {X.}~\bibnamefont {Zhao}}, \bibinfo {author} {\bibfnamefont {Y.}~\bibnamefont {Yang}},\ and\ \bibinfo {author} {\bibfnamefont {G.}~\bibnamefont {Chiribella}},\ }\bibfield  {title} {\bibinfo {title} {{Quantum Metrology with Indefinite Causal Order}},\ }\href {https://doi.org/10.1103/PhysRevLett.124.190503} {\bibfield  {journal} {\bibinfo  {journal} {Phys. Rev. Lett.}\ }\textbf {\bibinfo {volume} {124}},\ \bibinfo {pages} {190503} (\bibinfo {year} {2020})}\BibitemShut {NoStop}%
\bibitem [{\citenamefont {Loizeau}\ and\ \citenamefont {Grinbaum}(2020)}]{Loizeau2020}%
  \BibitemOpen
  \bibfield  {author} {\bibinfo {author} {\bibfnamefont {N.}~\bibnamefont {Loizeau}}\ and\ \bibinfo {author} {\bibfnamefont {A.}~\bibnamefont {Grinbaum}},\ }\bibfield  {title} {\bibinfo {title} {Channel capacity enhancement with indefinite causal order},\ }\href {https://doi.org/10.1103/PhysRevA.101.012340} {\bibfield  {journal} {\bibinfo  {journal} {Phys. Rev. A}\ }\textbf {\bibinfo {volume} {101}},\ \bibinfo {pages} {012340} (\bibinfo {year} {2020})}\BibitemShut {NoStop}%
\bibitem [{\citenamefont {Chiribella}\ \emph {et~al.}(2021{\natexlab{a}})\citenamefont {Chiribella}, \citenamefont {Banik}, \citenamefont {Bhattacharya}, \citenamefont {Guha}, \citenamefont {Alimuddin}, \citenamefont {Roy}, \citenamefont {Saha}, \citenamefont {Agrawal},\ and\ \citenamefont {Kar}}]{Chiribella2021}%
  \BibitemOpen
  \bibfield  {author} {\bibinfo {author} {\bibfnamefont {G.}~\bibnamefont {Chiribella}}, \bibinfo {author} {\bibfnamefont {M.}~\bibnamefont {Banik}}, \bibinfo {author} {\bibfnamefont {S.~S.}\ \bibnamefont {Bhattacharya}}, \bibinfo {author} {\bibfnamefont {T.}~\bibnamefont {Guha}}, \bibinfo {author} {\bibfnamefont {M.}~\bibnamefont {Alimuddin}}, \bibinfo {author} {\bibfnamefont {A.}~\bibnamefont {Roy}}, \bibinfo {author} {\bibfnamefont {S.}~\bibnamefont {Saha}}, \bibinfo {author} {\bibfnamefont {S.}~\bibnamefont {Agrawal}},\ and\ \bibinfo {author} {\bibfnamefont {G.}~\bibnamefont {Kar}},\ }\bibfield  {title} {\bibinfo {title} {Indefinite causal order enables perfect quantum communication with zero capacity channels},\ }\href {https://doi.org/10.1088/1367-2630/abe7a0} {\bibfield  {journal} {\bibinfo  {journal} {New J. Phys.}\ }\textbf {\bibinfo {volume} {23}},\ \bibinfo {pages} {033039} (\bibinfo {year} {2021}{\natexlab{a}})}\BibitemShut {NoStop}%
\bibitem [{\citenamefont {Chiribella}\ \emph {et~al.}(2013)\citenamefont {Chiribella}, \citenamefont {D'Ariano}, \citenamefont {Perinotti},\ and\ \citenamefont {Valiron}}]{Chiribella2013}%
  \BibitemOpen
  \bibfield  {author} {\bibinfo {author} {\bibfnamefont {G.}~\bibnamefont {Chiribella}}, \bibinfo {author} {\bibfnamefont {G.~M.}\ \bibnamefont {D'Ariano}}, \bibinfo {author} {\bibfnamefont {P.}~\bibnamefont {Perinotti}},\ and\ \bibinfo {author} {\bibfnamefont {B.}~\bibnamefont {Valiron}},\ }\bibfield  {title} {\bibinfo {title} {Quantum computations without definite causal structure},\ }\href {https://doi.org/10.1103/PhysRevA.88.022318} {\bibfield  {journal} {\bibinfo  {journal} {Phys. Rev. A}\ }\textbf {\bibinfo {volume} {88}},\ \bibinfo {pages} {022318} (\bibinfo {year} {2013})}\BibitemShut {NoStop}%
\bibitem [{\citenamefont {Oreshkov}\ \emph {et~al.}(2012)\citenamefont {Oreshkov}, \citenamefont {Costa},\ and\ \citenamefont {Brukner}}]{Oreshkov2012}%
  \BibitemOpen
  \bibfield  {author} {\bibinfo {author} {\bibfnamefont {O.}~\bibnamefont {Oreshkov}}, \bibinfo {author} {\bibfnamefont {F.}~\bibnamefont {Costa}},\ and\ \bibinfo {author} {\bibfnamefont {Ä.}~\bibnamefont {Brukner}},\ }\bibfield  {title} {\bibinfo {title} {Quantum correlations with no causal order},\ }\href {https://doi.org/10.1038/ncomms2076} {\bibfield  {journal} {\bibinfo  {journal} {Nat. Commun.}\ }\textbf {\bibinfo {volume} {3}},\ \bibinfo {pages} {1092} (\bibinfo {year} {2012})}\BibitemShut {NoStop}%
\bibitem [{\citenamefont {Ara\'ujo}\ \emph {et~al.}(2014)\citenamefont {Ara\'ujo}, \citenamefont {Costa},\ and\ \citenamefont {Brukner}}]{Araujo2014}%
  \BibitemOpen
  \bibfield  {author} {\bibinfo {author} {\bibfnamefont {M.}~\bibnamefont {Ara\'ujo}}, \bibinfo {author} {\bibfnamefont {F.}~\bibnamefont {Costa}},\ and\ \bibinfo {author} {\bibfnamefont {{\v{C}}.}~\bibnamefont {Brukner}},\ }\bibfield  {title} {\bibinfo {title} {{Computational Advantage from Quantum-Controlled Ordering of Gates}},\ }\href {https://doi.org/10.1103/PhysRevLett.113.250402} {\bibfield  {journal} {\bibinfo  {journal} {Phys. Rev. Lett.}\ }\textbf {\bibinfo {volume} {113}},\ \bibinfo {pages} {250402} (\bibinfo {year} {2014})}\BibitemShut {NoStop}%
\bibitem [{\citenamefont {Felce}\ and\ \citenamefont {Vedral}(2020)}]{Felce2020}%
  \BibitemOpen
  \bibfield  {author} {\bibinfo {author} {\bibfnamefont {D.}~\bibnamefont {Felce}}\ and\ \bibinfo {author} {\bibfnamefont {V.}~\bibnamefont {Vedral}},\ }\bibfield  {title} {\bibinfo {title} {{Quantum Refrigeration with Indefinite Causal Order}},\ }\href {https://doi.org/10.1103/PhysRevLett.125.070603} {\bibfield  {journal} {\bibinfo  {journal} {Phys. Rev. Lett.}\ }\textbf {\bibinfo {volume} {125}},\ \bibinfo {pages} {070603} (\bibinfo {year} {2020})}\BibitemShut {NoStop}%
\bibitem [{\citenamefont {Simonov}\ \emph {et~al.}(2022)\citenamefont {Simonov}, \citenamefont {Francica}, \citenamefont {Guarnieri},\ and\ \citenamefont {Paternostro}}]{Simonov2022}%
  \BibitemOpen
  \bibfield  {author} {\bibinfo {author} {\bibfnamefont {K.}~\bibnamefont {Simonov}}, \bibinfo {author} {\bibfnamefont {G.}~\bibnamefont {Francica}}, \bibinfo {author} {\bibfnamefont {G.}~\bibnamefont {Guarnieri}},\ and\ \bibinfo {author} {\bibfnamefont {M.}~\bibnamefont {Paternostro}},\ }\bibfield  {title} {\bibinfo {title} {Work extraction from coherently activated maps via quantum switch},\ }\href {https://doi.org/10.1103/PhysRevA.105.032217} {\bibfield  {journal} {\bibinfo  {journal} {Phys. Rev. A}\ }\textbf {\bibinfo {volume} {105}},\ \bibinfo {pages} {032217} (\bibinfo {year} {2022})}\BibitemShut {NoStop}%
\bibitem [{\citenamefont {Francica}(2022)}]{Francica2022-2}%
  \BibitemOpen
  \bibfield  {author} {\bibinfo {author} {\bibfnamefont {G.}~\bibnamefont {Francica}},\ }\bibfield  {title} {\bibinfo {title} {Causal games of work extraction with indefinite causal order},\ }\href {https://doi.org/10.1103/PhysRevA.106.042214} {\bibfield  {journal} {\bibinfo  {journal} {Phys. Rev. A}\ }\textbf {\bibinfo {volume} {106}},\ \bibinfo {pages} {042214} (\bibinfo {year} {2022})}\BibitemShut {NoStop}%
\bibitem [{\citenamefont {Procopio}\ \emph {et~al.}(2019)\citenamefont {Procopio}, \citenamefont {Delgado}, \citenamefont {Enr\'{i}quez}, \citenamefont {Belabas},\ and\ \citenamefont {Levenson}}]{Procopio2019}%
  \BibitemOpen
  \bibfield  {author} {\bibinfo {author} {\bibfnamefont {L.~M.}\ \bibnamefont {Procopio}}, \bibinfo {author} {\bibfnamefont {F.}~\bibnamefont {Delgado}}, \bibinfo {author} {\bibfnamefont {M.}~\bibnamefont {Enr\'{i}quez}}, \bibinfo {author} {\bibfnamefont {N.}~\bibnamefont {Belabas}},\ and\ \bibinfo {author} {\bibfnamefont {J.~A.}\ \bibnamefont {Levenson}},\ }\bibfield  {title} {\bibinfo {title} {{Communication Enhancement through Quantum Coherent Control of N Channels in an Indefinite Causal-Order Scenario}},\ }\href {https://doi.org/10.3390/e21101012} {\bibfield  {journal} {\bibinfo  {journal} {Entropy}\ }\textbf {\bibinfo {volume} {21}},\ \bibinfo {pages} {1012} (\bibinfo {year} {2019})}\BibitemShut {NoStop}%
\bibitem [{\citenamefont {Chiribella}\ \emph {et~al.}(2021{\natexlab{b}})\citenamefont {Chiribella}, \citenamefont {Wilson},\ and\ \citenamefont {Chau}}]{Chiribella2021-2}%
  \BibitemOpen
  \bibfield  {author} {\bibinfo {author} {\bibfnamefont {G.}~\bibnamefont {Chiribella}}, \bibinfo {author} {\bibfnamefont {M.}~\bibnamefont {Wilson}},\ and\ \bibinfo {author} {\bibfnamefont {H.~F.}\ \bibnamefont {Chau}},\ }\bibfield  {title} {\bibinfo {title} {{Quantum and Classical Data Transmission through Completely Depolarizing Channels in a Superposition of Cyclic Orders}},\ }\href {https://doi.org/10.1103/PhysRevLett.127.190502} {\bibfield  {journal} {\bibinfo  {journal} {Phys. Rev. Lett.}\ }\textbf {\bibinfo {volume} {127}},\ \bibinfo {pages} {190502} (\bibinfo {year} {2021}{\natexlab{b}})}\BibitemShut {NoStop}%
\bibitem [{\citenamefont {Wang}\ \emph {et~al.}(2025)\citenamefont {Wang}, \citenamefont {Zhou}, \citenamefont {Feng}, \citenamefont {Nie}, \citenamefont {Xia}, \citenamefont {Xiao}, \citenamefont {Li}, \citenamefont {Vedral},\ and\ \citenamefont {Zhou}}]{Wang2025}%
  \BibitemOpen
  \bibfield  {author} {\bibinfo {author} {\bibfnamefont {Y.}~\bibnamefont {Wang}}, \bibinfo {author} {\bibfnamefont {L.}~\bibnamefont {Zhou}}, \bibinfo {author} {\bibfnamefont {T.}~\bibnamefont {Feng}}, \bibinfo {author} {\bibfnamefont {H.}~\bibnamefont {Nie}}, \bibinfo {author} {\bibfnamefont {Y.}~\bibnamefont {Xia}}, \bibinfo {author} {\bibfnamefont {T.}~\bibnamefont {Xiao}}, \bibinfo {author} {\bibfnamefont {J.}~\bibnamefont {Li}}, \bibinfo {author} {\bibfnamefont {V.}~\bibnamefont {Vedral}},\ and\ \bibinfo {author} {\bibfnamefont {X.}~\bibnamefont {Zhou}},\ }\bibfield  {title} {\bibinfo {title} {Experimental demonstration of genuine quantum information transmission through completely depolarizing channels in a superposition of cyclic orders},\ }\href {https://arxiv.org/abs/2510.07127} {\bibfield  {journal} {\bibinfo  {journal} {arXiv:2510.07127}\ } (\bibinfo {year} {2025})}\BibitemShut {NoStop}%
\bibitem [{\citenamefont {Zhu}\ \emph {et~al.}(2023)\citenamefont {Zhu}, \citenamefont {Chen}, \citenamefont {Hasegawa},\ and\ \citenamefont {Xue}}]{Zhu2023}%
  \BibitemOpen
  \bibfield  {author} {\bibinfo {author} {\bibfnamefont {G.}~\bibnamefont {Zhu}}, \bibinfo {author} {\bibfnamefont {Y.}~\bibnamefont {Chen}}, \bibinfo {author} {\bibfnamefont {Y.}~\bibnamefont {Hasegawa}},\ and\ \bibinfo {author} {\bibfnamefont {P.}~\bibnamefont {Xue}},\ }\bibfield  {title} {\bibinfo {title} {{Charging Quantum Batteries via Indefinite Causal Order: Theory and Experiment}},\ }\href {https://doi.org/10.1103/PhysRevLett.131.240401} {\bibfield  {journal} {\bibinfo  {journal} {Phys. Rev. Lett.}\ }\textbf {\bibinfo {volume} {131}},\ \bibinfo {pages} {240401} (\bibinfo {year} {2023})}\BibitemShut {NoStop}%
\bibitem [{\citenamefont {Li}\ \emph {et~al.}(2025)\citenamefont {Li}, \citenamefont {Ma}, \citenamefont {Hao},\ and\ \citenamefont {Yu}}]{Li2025}%
  \BibitemOpen
  \bibfield  {author} {\bibinfo {author} {\bibfnamefont {H.}~\bibnamefont {Li}}, \bibinfo {author} {\bibfnamefont {H.}~\bibnamefont {Ma}}, \bibinfo {author} {\bibfnamefont {Y.}~\bibnamefont {Hao}},\ and\ \bibinfo {author} {\bibfnamefont {W.}~\bibnamefont {Yu}},\ }\bibfield  {title} {\bibinfo {title} {Enhancing ergotropy of a quantum battery with coherent chargers: The catalystlike role of indefinite causal order},\ }\href {https://doi.org/10.1103/s6dl-zgkx} {\bibfield  {journal} {\bibinfo  {journal} {Phys. Rev. A}\ }\textbf {\bibinfo {volume} {112}},\ \bibinfo {pages} {042229} (\bibinfo {year} {2025})}\BibitemShut {NoStop}%
\bibitem [{\citenamefont {Biswas}\ \emph {et~al.}(2022)\citenamefont {Biswas}, \citenamefont {{\L{}}obejko}, \citenamefont {Mazurek}, \citenamefont {Ja{\l{}}owiecki},\ and\ \citenamefont {Horodecki}}]{Biswas2022}%
  \BibitemOpen
  \bibfield  {author} {\bibinfo {author} {\bibfnamefont {T.}~\bibnamefont {Biswas}}, \bibinfo {author} {\bibfnamefont {M.}~\bibnamefont {{\L{}}obejko}}, \bibinfo {author} {\bibfnamefont {P.}~\bibnamefont {Mazurek}}, \bibinfo {author} {\bibfnamefont {K.}~\bibnamefont {Ja{\l{}}owiecki}},\ and\ \bibinfo {author} {\bibfnamefont {M.}~\bibnamefont {Horodecki}},\ }\bibfield  {title} {\bibinfo {title} {Extraction of ergotropy: free energy bound and application to open cycle engines},\ }\href {https://doi.org/10.22331/q-2022-10-17-841} {\bibfield  {journal} {\bibinfo  {journal} {{Quantum}}\ }\textbf {\bibinfo {volume} {6}},\ \bibinfo {pages} {841} (\bibinfo {year} {2022})}\BibitemShut {NoStop}%
\bibitem [{\citenamefont {Song}\ \emph {et~al.}(2024{\natexlab{a}})\citenamefont {Song}, \citenamefont {Song}, \citenamefont {Ye},\ and\ \citenamefont {Wang}}]{MLSong2024}%
  \BibitemOpen
  \bibfield  {author} {\bibinfo {author} {\bibfnamefont {M.-L.}\ \bibnamefont {Song}}, \bibinfo {author} {\bibfnamefont {X.-K.}\ \bibnamefont {Song}}, \bibinfo {author} {\bibfnamefont {L.}~\bibnamefont {Ye}},\ and\ \bibinfo {author} {\bibfnamefont {D.}~\bibnamefont {Wang}},\ }\bibfield  {title} {\bibinfo {title} {Evaluating extractable work of quantum batteries via entropic uncertainty relations},\ }\href {https://doi.org/10.1103/PhysRevE.109.064103} {\bibfield  {journal} {\bibinfo  {journal} {Phys. Rev. E}\ }\textbf {\bibinfo {volume} {109}},\ \bibinfo {pages} {064103} (\bibinfo {year} {2024}{\natexlab{a}})}\BibitemShut {NoStop}%
\bibitem [{\citenamefont {Preskill}(2018)}]{Preskill2018}%
  \BibitemOpen
  \bibfield  {author} {\bibinfo {author} {\bibfnamefont {J.}~\bibnamefont {Preskill}},\ }\bibfield  {title} {\bibinfo {title} {{Quantum Computing in the NISQ era and beyond}},\ }\href {https://doi.org/10.22331/q-2018-08-06-79} {\bibfield  {journal} {\bibinfo  {journal} {Quantum}\ }\textbf {\bibinfo {volume} {2}},\ \bibinfo {pages} {79} (\bibinfo {year} {2018})}\BibitemShut {NoStop}%
\bibitem [{\citenamefont {Bharti}\ \emph {et~al.}(2022)\citenamefont {Bharti}, \citenamefont {Cervera-Lierta}, \citenamefont {Kyaw}, \citenamefont {Haug}, \citenamefont {Alperin-Lea}, \citenamefont {Anand}, \citenamefont {Degroote}, \citenamefont {Heimonen}, \citenamefont {Kottmann}, \citenamefont {Menke}, \citenamefont {Mok}, \citenamefont {Sim}, \citenamefont {Kwek},\ and\ \citenamefont {Aspuru-Guzik}}]{Bharti2022}%
  \BibitemOpen
  \bibfield  {author} {\bibinfo {author} {\bibfnamefont {K.}~\bibnamefont {Bharti}}, \bibinfo {author} {\bibfnamefont {A.}~\bibnamefont {Cervera-Lierta}}, \bibinfo {author} {\bibfnamefont {T.~H.}\ \bibnamefont {Kyaw}}, \bibinfo {author} {\bibfnamefont {T.}~\bibnamefont {Haug}}, \bibinfo {author} {\bibfnamefont {S.}~\bibnamefont {Alperin-Lea}}, \bibinfo {author} {\bibfnamefont {A.}~\bibnamefont {Anand}}, \bibinfo {author} {\bibfnamefont {M.}~\bibnamefont {Degroote}}, \bibinfo {author} {\bibfnamefont {H.}~\bibnamefont {Heimonen}}, \bibinfo {author} {\bibfnamefont {J.~S.}\ \bibnamefont {Kottmann}}, \bibinfo {author} {\bibfnamefont {T.}~\bibnamefont {Menke}}, \bibinfo {author} {\bibfnamefont {W.-K.}\ \bibnamefont {Mok}}, \bibinfo {author} {\bibfnamefont {S.}~\bibnamefont {Sim}}, \bibinfo {author} {\bibfnamefont {L.-C.}\ \bibnamefont {Kwek}},\ and\ \bibinfo {author} {\bibfnamefont {A.}~\bibnamefont {Aspuru-Guzik}},\ }\bibfield  {title} {\bibinfo {title} {Noisy intermediate-scale quantum algorithms},\ }\href
  {https://doi.org/10.1103/RevModPhys.94.015004} {\bibfield  {journal} {\bibinfo  {journal} {Rev. Mod. Phys.}\ }\textbf {\bibinfo {volume} {94}},\ \bibinfo {pages} {015004} (\bibinfo {year} {2022})}\BibitemShut {NoStop}%
\bibitem [{\citenamefont {Magesan}\ \emph {et~al.}(2011)\citenamefont {Magesan}, \citenamefont {Gambetta},\ and\ \citenamefont {Emerson}}]{Magesan2011}%
  \BibitemOpen
  \bibfield  {author} {\bibinfo {author} {\bibfnamefont {E.}~\bibnamefont {Magesan}}, \bibinfo {author} {\bibfnamefont {J.~M.}\ \bibnamefont {Gambetta}},\ and\ \bibinfo {author} {\bibfnamefont {J.}~\bibnamefont {Emerson}},\ }\bibfield  {title} {\bibinfo {title} {{Scalable and Robust Randomized Benchmarking of Quantum Processes}},\ }\href {https://doi.org/10.1103/PhysRevLett.106.180504} {\bibfield  {journal} {\bibinfo  {journal} {Phys. Rev. Lett.}\ }\textbf {\bibinfo {volume} {106}},\ \bibinfo {pages} {180504} (\bibinfo {year} {2011})}\BibitemShut {NoStop}%
\bibitem [{\citenamefont {Song}\ \emph {et~al.}(2024{\natexlab{b}})\citenamefont {Song}, \citenamefont {Liu}, \citenamefont {Zhou}, \citenamefont {Yang},\ and\ \citenamefont {An}}]{Song2024}%
  \BibitemOpen
  \bibfield  {author} {\bibinfo {author} {\bibfnamefont {W.-L.}\ \bibnamefont {Song}}, \bibinfo {author} {\bibfnamefont {H.-B.}\ \bibnamefont {Liu}}, \bibinfo {author} {\bibfnamefont {B.}~\bibnamefont {Zhou}}, \bibinfo {author} {\bibfnamefont {W.-L.}\ \bibnamefont {Yang}},\ and\ \bibinfo {author} {\bibfnamefont {J.-H.}\ \bibnamefont {An}},\ }\bibfield  {title} {\bibinfo {title} {{Remote Charging and Degradation Suppression for the Quantum Battery}},\ }\href {https://doi.org/10.1103/PhysRevLett.132.090401} {\bibfield  {journal} {\bibinfo  {journal} {Phys. Rev. Lett.}\ }\textbf {\bibinfo {volume} {132}},\ \bibinfo {pages} {090401} (\bibinfo {year} {2024}{\natexlab{b}})}\BibitemShut {NoStop}%
\bibitem [{\citenamefont {Tirone}\ \emph {et~al.}(2025)\citenamefont {Tirone}, \citenamefont {Andolina}, \citenamefont {Calaj\`o}, \citenamefont {Giovannetti},\ and\ \citenamefont {Rossini}}]{Tirone2025}%
  \BibitemOpen
  \bibfield  {author} {\bibinfo {author} {\bibfnamefont {S.}~\bibnamefont {Tirone}}, \bibinfo {author} {\bibfnamefont {G.~M.}\ \bibnamefont {Andolina}}, \bibinfo {author} {\bibfnamefont {G.}~\bibnamefont {Calaj\`o}}, \bibinfo {author} {\bibfnamefont {V.}~\bibnamefont {Giovannetti}},\ and\ \bibinfo {author} {\bibfnamefont {D.}~\bibnamefont {Rossini}},\ }\bibfield  {title} {\bibinfo {title} {Many-body enhancement of energy storage in a waveguide {QED} quantum battery},\ }\href {https://doi.org/10.1103/c4hd-8vfj} {\bibfield  {journal} {\bibinfo  {journal} {Phys. Rev. A}\ }\textbf {\bibinfo {volume} {112}},\ \bibinfo {pages} {013717} (\bibinfo {year} {2025})}\BibitemShut {NoStop}%
\bibitem [{\citenamefont {Guo}\ \emph {et~al.}(2025)\citenamefont {Guo}, \citenamefont {Cao},\ and\ \citenamefont {Zhao}}]{Guo2025}%
  \BibitemOpen
  \bibfield  {author} {\bibinfo {author} {\bibfnamefont {Y.}~\bibnamefont {Guo}}, \bibinfo {author} {\bibfnamefont {L.}~\bibnamefont {Cao}},\ and\ \bibinfo {author} {\bibfnamefont {J.}~\bibnamefont {Zhao}},\ }\bibfield  {title} {\bibinfo {title} {Nonreciprocal open quantum battery network in a photonic waveguide array},\ }\href {https://doi.org/10.1103/fn1b-2m9g} {\bibfield  {journal} {\bibinfo  {journal} {Phys. Rev. A}\ }\textbf {\bibinfo {volume} {111}},\ \bibinfo {pages} {063520} (\bibinfo {year} {2025})}\BibitemShut {NoStop}%
\bibitem [{\citenamefont {Salvia}\ \emph {et~al.}(2023)\citenamefont {Salvia}, \citenamefont {Perarnau-Llobet}, \citenamefont {Haack}, \citenamefont {Brunner},\ and\ \citenamefont {Nimmrichter}}]{Salvia2023}%
  \BibitemOpen
  \bibfield  {author} {\bibinfo {author} {\bibfnamefont {R.}~\bibnamefont {Salvia}}, \bibinfo {author} {\bibfnamefont {M.}~\bibnamefont {Perarnau-Llobet}}, \bibinfo {author} {\bibfnamefont {G.}~\bibnamefont {Haack}}, \bibinfo {author} {\bibfnamefont {N.}~\bibnamefont {Brunner}},\ and\ \bibinfo {author} {\bibfnamefont {S.}~\bibnamefont {Nimmrichter}},\ }\bibfield  {title} {\bibinfo {title} {Quantum advantage in charging cavity and spin batteries by repeated interactions},\ }\href {https://doi.org/10.1103/PhysRevResearch.5.013155} {\bibfield  {journal} {\bibinfo  {journal} {Phys. Rev. Res.}\ }\textbf {\bibinfo {volume} {5}},\ \bibinfo {pages} {013155} (\bibinfo {year} {2023})}\BibitemShut {NoStop}%
\bibitem [{\citenamefont {Rinaldi}\ \emph {et~al.}(2025)\citenamefont {Rinaldi}, \citenamefont {Filip}, \citenamefont {Gerace},\ and\ \citenamefont {Guarnieri}}]{Rinaldi2025}%
  \BibitemOpen
  \bibfield  {author} {\bibinfo {author} {\bibfnamefont {D.}~\bibnamefont {Rinaldi}}, \bibinfo {author} {\bibfnamefont {R.}~\bibnamefont {Filip}}, \bibinfo {author} {\bibfnamefont {D.}~\bibnamefont {Gerace}},\ and\ \bibinfo {author} {\bibfnamefont {G.}~\bibnamefont {Guarnieri}},\ }\bibfield  {title} {\bibinfo {title} {Reliable quantum advantage in quantum battery charging},\ }\href {https://doi.org/10.1103/6kwv-z6fx} {\bibfield  {journal} {\bibinfo  {journal} {Phys. Rev. A}\ }\textbf {\bibinfo {volume} {112}},\ \bibinfo {pages} {012205} (\bibinfo {year} {2025})}\BibitemShut {NoStop}%
\bibitem [{\citenamefont {Morrone}\ \emph {et~al.}(2023)\citenamefont {Morrone}, \citenamefont {Rossi}, \citenamefont {Smirne},\ and\ \citenamefont {Genoni}}]{Morrone2023}%
  \BibitemOpen
  \bibfield  {author} {\bibinfo {author} {\bibfnamefont {D.}~\bibnamefont {Morrone}}, \bibinfo {author} {\bibfnamefont {M.~A.~C.}\ \bibnamefont {Rossi}}, \bibinfo {author} {\bibfnamefont {A.}~\bibnamefont {Smirne}},\ and\ \bibinfo {author} {\bibfnamefont {M.~G.}\ \bibnamefont {Genoni}},\ }\bibfield  {title} {\bibinfo {title} {Charging a quantum battery in a non-{M}arkovian environment: a collisional model approach},\ }\href {https://doi.org/10.1088/2058-9565/accca4} {\bibfield  {journal} {\bibinfo  {journal} {QST}\ }\textbf {\bibinfo {volume} {8}},\ \bibinfo {pages} {035007} (\bibinfo {year} {2023})}\BibitemShut {NoStop}%
\end{thebibliography}

\end{document}